\documentclass[journal,10pt]{IEEEtran}
\usepackage{moreverb}
\usepackage{epsfig}
\usepackage{amsmath,amssymb,amsthm,mathrsfs,amsfonts,dsfont}
\usepackage{adjustbox,lipsum}
\usepackage[linesnumbered,ruled,vlined]{algorithm2e}
\usepackage{amsfonts}
\usepackage{epsfig}
\usepackage{amssymb}
\usepackage{amsmath}
\usepackage{amsthm}
\usepackage{multirow}
\usepackage[font={small}]{caption}

\usepackage{booktabs}
\usepackage{makecell}  

\usepackage{setspace}
\usepackage{rotating}
\usepackage{graphicx}
\usepackage{tabularx}
\usepackage{array}
\usepackage{anyfontsize}
\usepackage{color,soul}
\usepackage{graphicx,dblfloatfix}
\usepackage{epstopdf}
\usepackage{blindtext}
\usepackage{amsmath}
\usepackage{amsthm,amssymb,amsmath,bm}
\usepackage{subfigure}
\usepackage{amsfonts}
\usepackage{epsfig}
\usepackage{amssymb}
\usepackage{amsmath}
\usepackage{cite}
\usepackage{graphicx}
\usepackage{fancyhdr}
\usepackage[font={small}]{caption}
\usepackage{tabularx}
\usepackage{tcolorbox}
\usepackage{cite}
\usepackage{setspace}
\usepackage{flushend}
\usepackage[inline]{enumitem}

\newtheorem{theorem}{Theorem}
\newtheorem{lemma}{Lemma}

\newtheorem{example}{Example}
\newtheorem{definition}{Definition}
\newtheorem{remark}{Remark}
\newtheorem{proposition}{Proposition}

\allowdisplaybreaks
\title{{An Unconditionally Secure Encryption Scheme for IoBT Networks}} 

\author{
\IEEEauthorblockN{Mohammad~Moltafet, Hamid~R.~Sadjadpour, and Zouheir~Rezki
}
\thanks{The authors are with the Department of Electrical and Computer Engineering, University of California Santa Cruz (UCSC), Santa Cruz, CA 95064, USA (e-mail: mmoltafe@ucsc.edu, hamid@ucsc.edu, zrezki@ucsc.edu).
}
}  
\begin{document}
\maketitle
\sloppy

\begin{abstract}
We consider an Internet of Battlefield Things (IoBT) system consisting of multiple devices that want to securely communicate with each other during a mission in the presence of an adversary with unbounded computational power. The adversary has complete access to listen/read the ciphertext without tampering with the communication line. We provide an unconditionally secure encryption scheme to exchange messages among devices in the system. The main idea behind the scheme is to provide secret keys to exchange messages using a random binary matrix that is securely shared among all the devices, and pair-wise random secret keys established between each pair of devices attempting to communicate before the mission. The scheme is implemented by using finite group modular addition. We show that the scheme is \textit{absolutely} semantically secure, i.e., the scheme guarantees that an adversary with unbounded computational power cannot get even one bit of information about a message, except for an exponentially small probability in a security parameter. Besides that, we show that even if the random binary matrix is revealed to the adversary, the provided scheme is computationally secure against the key recovery attack. 

\end{abstract}

\section{Introduction}
The most effective approach of secrecy coding that provides perfect secrecy is the one-time pad (or Vernam cipher). Vernam introduced his cipher in 1926 \cite{Vernam1926}, and Shannon in 1949 \cite{Shannon} proved that the one-time pad scheme provides perfect secrecy. According to the one-time pad scheme, to securely exchange a message, an independent and uniformly distributed one-time secret key whose size equals the message length needs to be established between the transmitter and receiver. More specifically, in \cite{Shannon}, Shannon proved that perfect secrecy against an all-powerful adversary with unbounded storage capacity and unbounded computational power who has complete access to the communication line is only achievable if the uncertainty
of the secret key is at least as great as that of the plaintext, this is also known as {\it Shannon impossibility result}\cite{Yevgeniy2012}. 

In 1992, Maurer introduced the bounded storage model and proposed the first scheme in the bounded storage model that provides \textit{unconditional} secrecy in the seminal work \cite{Maurer}. In the bounded storage model, the adversary is computationally unbounded and has a bounded storage capacity.  
 In this model, unconditional secrecy is guaranteed by using a publicly available random string whose length is much larger than the adversary's storage capacity. In this model, the adversary performs an attack in two phases. In phase one, first, the transmitter and receiver establish a secret key, and then, the random string is broadcast. Using the shared secret key, an encryption scheme, and the random string, the transmitter and receiver compute a final key to encrypt and decrypt the message. In this phase, the adversary can compute a function on the random string and store the result. In the second phase, the random string is not available and the adversary is provided with the ciphertext, the \textit{secret key}, and \textit{unbounded storage capacity} and unbounded computational power. He tries to get information about the encrypted message using the provided information. 
 In \cite{Maurer}, Maurer proved that by using his proposed scheme, the adversary cannot gain any information about the plaintext with a probability arbitrarily close to one.
 Following the introduction of the bounded storage model in  \cite{Maurer}, extensive research has been conducted to enhance its practicality. For instance, efforts have been made to enable the adversary to compute arbitrary functions on the publicly available random string \cite{Maurergeneralb,AumannZongRabin02}, as well as to reduce the size of the initial secret key and the length of the public random string \cite{AumannRabinAR,DziembowskiMaurer02,MMTIFS2024}. However, the main problem with the bounded storage model is that it may not be an accurate representation of real-world adversaries. In practice, adversaries may have access to sufficient storage, making the bounded storage model vulnerable.
 %
 %
 %
%

The security of many commonly used cryptographic schemes, both asymmetric and symmetric, is based on the assumption that the adversary/attacker has limited computational power. These schemes are said to provide \textit{computational security}, as they rely on the assumption that it would take an impractical amount of time or resources for an attacker to break the encryption scheme. 
For example, the security of the well-known  Rivest-Shamir–Adleman (RSA) public-key cryptosystem \cite{rivest1978method} is based on the (unproven) hardness of factoring large numbers, and the security of the  Diffie-Hellman protocol \cite{mngdifi1976} is based on the (unproven) hardness of computing discrete logarithms in certain groups.
%
Despite their widespread use, one major drawback of protocols that provide computational security is that they do not guarantee \textit{everlasting security} for top-secret messages. This means that an adversary could potentially store the ciphertext at the time of transmission and decipher it later, when advances in computing technology and code-breaking algorithms make it possible. As a result, there is a need for cryptographic protocols that provide stronger guarantees of security, which can withstand attacks from powerful adversaries even in the long term.

 In 1994, Peter Shor developed a polynomial-time quantum algorithm for factoring integers and solving discrete logarithm problems \cite{Shor}. Thus, the deployment of quantum computers in the future will render many existing encryption protocols vulnerable \cite{QuantumComputers}, e.g., the currently popular public-key cryptosystems, such as RSA and Elliptic Curve Cryptography (ECC), will be broken in the era of quantum computers. Code-based public-key cryptosystems which are based on the hardness of decoding a random linear code, such as Goppa codes (used in the original code-based  public-key scheme developed by Robert McEliece) \cite{mceliece1978public}, Generalized Reed-Solomon codes \cite{Baldi,Berger,Bolkema,Khathuria}, 
Reed-Muller codes \cite{Vladimir}, algebraic-geometric codes \cite{Janwa}, and Moderate-Density Parity Check and Low-Density Parity Check codes \cite{Misoczki,Bodrato}, are considered promising approaches for the post-quantum era.
However, code-based cryptosystems have some practical limitations, including larger key sizes 
compared to other public key cryptosystems, and they could be vulnerable to structural attacks \cite{Couvreur,Pellikaan,Otmani,Otmani2,PolyGaborit2018}. It is worth emphasizing that code-based cryptosystems provide computational security and therefore do not guarantee everlasting security.

Like all computationally secure cryptographic systems, symmetric cryptography systems are vulnerable to attacks over time. The Data Encryption Standard (DES), which was previously the official Federal Information Processing Standard (FIPS) in the United States, is no longer considered secure, as it can be compromised within a relatively short time, usually a few hours. 
Currently, the widely used and recommended symmetric encryption algorithm is the Advanced Encryption Standard (AES). AES is considered a secure cryptosystem, at least with the present technology. Nevertheless, there is always the possibility of future advancements in technology that may render AES vulnerable to attacks. For example, quantum computers can significantly affect the security level of symmetric cryptosystems. In Table~\ref{Tabelsym}, the security level of some of the currently used symmetric encryption algorithms against the quantum search algorithm attack (discovered by Grover \cite{grover1996fast}) is presented \cite{bernstein2017post}. As can be seen, the security level of the algorithms decreases by a factor of $2$ compared to that against pre-quantum attacks.
\begin{table*}
\centering
\caption{The security level of some of the currently used symmetric encryption algorithms against the quantum search algorithm attack.}\label{Tabelsym}
\begin{tabular}{|l|l|c|c|}
\hline
\textbf{Name} & \textbf{Function} & \multicolumn{1}{p{2.5cm}|}{\centering \textbf{Pre-quantum security level}} & \multicolumn{1}{p{2.5cm}|}{\centering \textbf{Post-quantum security level}} \\ \hline
AES-128  & Block cipher & 128 & 64  \\ \hline
AES-256  & Block cipher & 256 & 128  \\ \hline
Salsa20  & Stream cipher & 256 & 128  \\ \hline
SHA-256  & Hash function & 256 & 128  \\ \hline
SHA-3  & Hash function & 256 & 128  \\ \hline
\end{tabular}
\end{table*}

The works in \cite{9372286,8731758,9695397,9250428,9825730} studied \textit{computationally secure} encryption schemes for  (low-power) sensor networks, e.g., Internet of Battlefield Things (IoBT).  
A comprehensive literature review of symmetric and asymmetric encryption schemes for sensor networks
can be found in \cite{pandey2024recent}. It is worth noting that all the discussed schemes in \cite{pandey2024recent} provide computational security.
%
%
%
In contrast to schemes that rely on computational assumptions, unconditionally secure schemes guarantee that the adversary cannot extract any information about the plaintext from the ciphertext, regardless of their computational power. This property makes unconditionally secure schemes particularly appealing for scenarios where long-term security is essential, such as military communications, and storing a massive amount of sensitive data, e.g., financial transactions, and health-related data, on public cloud storage. 
In mission-based military applications, tactical and highly sensitive information is exchanged among devices. The development of quantum computers in the near future poses a serious threat to existing, commonly used encryption schemes, which could be broken by powerful quantum algorithms. Consequently, in this paper, we develop an encryption scheme that, regardless of adversaries’ computational power, guarantees secure communication between devices where the level of security can be adjusted via a security parameter. This aligns with the goals of \emph{post‑quantum cryptography}, providing protection even against future advances. Next, we present the main results of the paper.

\subsection{Our Results}
In this paper, we consider an IoBT system consisting of multiple devices communicating with each other in the presence of an all-powerful adversary with unbounded storage and unbounded computational power who has complete access to the communication line, i.e., 
 the adversary has complete access to listen/read the ciphertext without tampering with the communication line.
 Our main goal is to provide a scheme that is unconditionally secure against the adversary, i.e., regardless of the computational power of the adversary, the probability that the adversary can get even one bit of information about the plaintext is negligibly small. Inspired by the concept and the protocol presented for the bounded storage model in \cite{Maurer}, we provide a technique that achieves unconditional secrecy against the adversary. Unlike the bounded storage model, 
we provide unconditional security against an adversary with unbounded storage and unbounded computational power.
The main idea behind the scheme is to  
establish secret keys for the encryption and decryption of exchanged messages between each pair of devices. 
The scheme is implemented using a binary random matrix securely shared among all devices, along with pairwise secret keys established between each device pair before the mission.

The security we obtain for the provided encryption scheme is the \textit{absolute} version of the \textit{semantic security} introduced in \cite{Probabilistic1984Goldwasser}. It is the absolute version of semantic security because the scheme allows the adversary to have unbounded computational power \cite{AumannZongRabin02}.  A cryptographic scheme is called unconditionally secure if it is absolutely semantically secure \cite[Sect.~2.2]{8584398}, \cite{AumannZongRabin02,AumannRabinAR,DziembowskiMaurer02,DingRabin02}. Thus, the scheme provides everlasting security in contrast to commonly used protocols such as the RSA cryptosystem, Diffie-Hellman, and AES, which provide computational security.
The proposed scheme is easy to deploy on devices, i.e., it is implemented by exploiting finite group modular addition. Using the scheme, each device can securely exchange multiple messages with other devices in the system using the same random matrix, and the pair-wise secret keys are established once to securely exchange multiple messages.

We define the secrecy gain of the provided scheme as the ratio of the amount of data that a device can securely exchange with other devices and the number of secret bits required to be stored on the device side. We show that by choosing an appropriate value for the security parameter, one can significantly increase the number of devices that a specific device can securely communicate with, which in turn, increases the security gain.

In addition, we study the security of the scheme in a scenario where the binary random matrix is exposed to an adversary with bounded computational power; we show that even if the binary random matrix is revealed to the adversary, the provided scheme is computationally secure against the key recovery attack. 


To the best of our knowledge, this is the first work to provide an unconditionally secure encryption scheme for the considered system. 

Preliminary results of this work were presented in \cite{milcome2023}. The main differences between the current version and \cite{milcome2023} are as follows:
(i) detailed proofs of the main theorems are included in this version; these were omitted in \cite{milcome2023} due to space constraints; (ii) a new section has been added that presents numerical results and performance evaluations, along with a discussion on potential future research directions; (iii) the overall presentation has been improved, for example, by introducing formal definitions of the underlying security notions; and
(iv) the introduction has been significantly expanded to provide a more comprehensive literature review and to better motivate the proposed scheme.

 \subsection{Organization}
The paper is organized as follows. The encryption and decryption scheme, along with the main results of the paper, are presented in Section~\ref{Protocol and Main Results}. The secrecy analysis is presented in Section~\ref {Multi-user Systems}.  In Section~\ref{Discussion}, numerical examples to assess the scheme and potential directions for future research are provided. Finally, concluding remarks are made in Section~\ref{Conclusions}.

\subsection{Notation}  
A random vector is denoted by a bold capital letter,  whereas the corresponding bold lowercase letter denotes a realization of the random vector, and a random variable is denoted by a capital letter, whereas the corresponding lowercase letter denotes a realization of the random variable.
Let $\mathcal{G}$ be a finite set$/$tuple, then, $G{\leftarrow} \mathcal{G}$ denotes choosing $G$ uniformly at random from $\mathcal{G}$. All the logarithm functions in this paper have base $2$.

\section{The Scheme and Summary of the Main Results}\label{Protocol and Main Results}
 \subsection{System Model}
\begin{figure}[t]
\centering
\includegraphics[width=1.05\linewidth,trim = 0mm 0mm 0mm 0mm,clip]{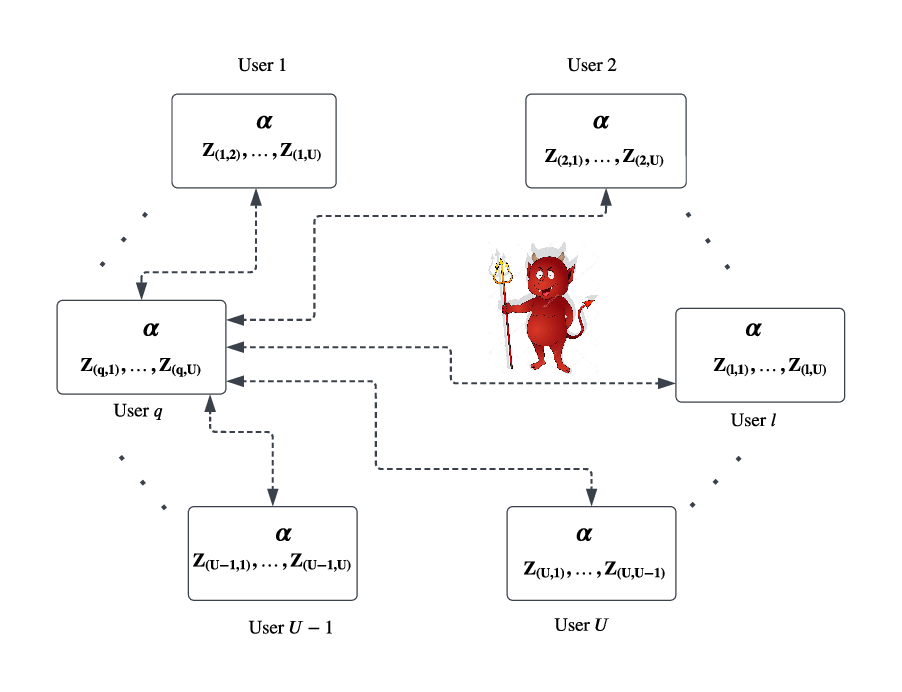}
\vspace{-6mm}
\caption{The considered system model.}
\vspace{-5mm}
\label{Model}
\end{figure}

We consider an IoBT system consisting of a set $\mathcal{U}=\{1,\ldots,U\}$ of $U$ devices and an adversary with 
unbounded computational power that has complete access to the communication line, i.e., the adversary can access ciphertexts transmitted from one device to another. The devices want to exchange highly confidential messages with each other during a mission in the presence of the adversary who acts as a passive eavesdropper, i.e., the adversary tries to get information from the exchanged messages between devices without tampering$/$jamming the communication link.
%
The set of all pairs of devices in the system is shown as $\mathcal{Q}=\{(1,2),\ldots,(1,U),(2,3),\ldots,(2,U),\ldots,(U-1,U)\}$.\footnote{Note that both $(q,l)$ and $(l,q)$ denote the pair of devices $q$ and $l$. Therefore, in indices representing the pair of devices, we have $(q,l)=(l,q)$.} 
 A typical illustration of the considered system model is shown in Fig.~\ref{Model}. 
The main goal is to develop an encryption scheme that ensures unconditional security for message exchange among the devices.



\subsection{The Encryption Scheme}
Let $k$ denote the system security parameter and  $\boldsymbol{\alpha}$ denote a $k\times n$ random binary matrix where 
$\boldsymbol{\alpha}[j,j']$ denotes the entry in the $j$th row and $j'$th column with $1\le j \le k$ and $0\le j' \le n-1$.
We assume that elements of the binary random matrix $\boldsymbol\alpha$ are uniformly distributed and statistically independent. Let $\mathbb{Z}_n=\{0,1,\ldots,n-1\}$ denote a group with addition modulo $n$ as the group operation, which is shown by $+_n$.

Prior to field operation, all 
$U$ devices are provisioned with the shared random binary matrix $\boldsymbol{\alpha}$, and pair-wise secret keys are established.
More specifically, devices $q$ and $l$ establish the pair-wise random secret key $\bold{Z}_{(q,l)}=({Z_{(q,l),1},\ldots,Z_{(q,l),k})}$
such that $\bold{Z}_{(q,l)}{\leftarrow} \mathbb{Z}_n^k$, where $\mathbb{Z}_n^k$ is the set of all $k$-tuples over $\mathbb{Z}_n$, thus of size $|\bold{Z}_{(q,l)}|=k\log n$ bits, and independent of all other pair-wise secret keys.

Now, suppose that device $q$ wants to send message ${{\bold{M}_{(q,l)}=({M_{(q,l),1},\ldots,M_{(q,l),m}})\in\{0,1\}^{m}}}$, with $k<m<\lfloor\dfrac{n+1}{2}\rfloor$, where $\lfloor.\rfloor$ is the floor function, to device $l$ during the mission.
{The main goal is to provide the final key ${\bold{X}_{(q,l)}=({X_{(q,l),1}},\ldots,X_{(q,l),m})\in\{0,1\}^{m}}$ for encryption and decryption of the message $\bold{M}_{(q,l)}$, i.e., device $q$ uses $\bold{X}_{(q,l)}$ for encryption and device $l$ uses $\bold{X}_{(q,l)}$ for decryption, guaranteeing the security of the message against the adversary who has access to the ciphertext}. 
To this end, each device, $q$ and $l$, forming the pair $(q,l)$, computes the $i$th bit of the final key $\bold{X}{(q,l)}$, denoted as $X{(q,l),i}$ for $i=1,\ldots,m$, using the shared secret key $\bold{Z}_{(q,l)}$ and the shared binary random matrix $\boldsymbol{\alpha}$ as follows:
%
%
%
%
 %
\begin{align}\label{eq2ql}
X_{(q,l),i}=\bigoplus_{j=1}^k\boldsymbol{\alpha}[j, Z_{(q,l),j}+_n{(i-1)}].
\end{align}
Now, device $q$ computes $\bold{C}_{(q,l)}=\bold{M}_{(q,l)}\oplus \bold{X}_{(q,l)}$, where $\oplus$ denotes bit-wise XOR, and sends $\bold{C}_{(q,l)}$ to device $l$ as the ciphertext, and device $l$ decrypts the message by computing $\bold{M}_{(q,l)}=\bold{C}_{(q,l)}\oplus \bold{X}_{(q,l)}$. 
The encryption and decryption steps are summarized in Algorithm~\ref{Alg_Prot}.

\begin{algorithm}[t]
\SetKwComment{Comment}{/*}{ }
\SetAlgoLined
Message: $\bold{M}_{(q,l)}=(M_{(q,l),1},\ldots,M_{(q,l),m})\in \{0,1\}^m$,

Secret key: $\bold{Z}_{(q,l)}=(Z_{(q,l),1},\ldots,Z_{(q,l),k})$
such that $\bold{Z}_{(q,l)}{\leftarrow} \mathbb{Z}_n^k$,

Shared~random~matrix:~$\boldsymbol\alpha$, ~~~~~~~~~~~~~~~~~~~~~~~~
     {\Comment{Steps 4-7 are executed separately on both device $q$ and device $l$ sides to generate the same key on both sides}}   
\For  {$i=1$ to $m$}{
Compute the $i$th bit of the final key $\bold{X}_{(q,l)}$, $X_{(q,l),i}$:

$X_{(q,l),i}
=\bigoplus_{j=1}^k\boldsymbol\alpha[j,Z_{(q,l),j}+_n(i-1)]
$,

}

Devices $q$ and $l$ set the final key $\bold{X}_{(q,l)}=(X_{(q,l),1},\ldots,X_{(q,l),m})$,

Device $q$ encrypts $\bold{C}_{(q,l)}=\bold{M}_{(q,l)}\oplus \bold{X}_{(q,l)}$, and sends $\bold{C}_{(q,l)}$ to device $l$,

Device $l$ decrypts  the message $\bold{M}_{(q,l)}=\bold{C}_{(q,l)}\oplus \bold{X}_{(q,l)}$.

\caption{Encryption and decryption protocol of the single message $\bold{M}_{(q,l)}$ transmitted from device $q$ to device $l$}
\label{Alg_Prot}
\end{algorithm} 
%

\subsection{Security Notion}
Prior to delving into the formal definition of the security notion, we first specify the concept of a negligible function.

\begin{definition}[Negligible function \cite{Boneh2020AppliedCryptography}]
    A function $\nu:\mathbb{N}\rightarrow\mathbb{R}$, where  $\mathbb{N}$ is the set of natural numbers and $\mathbb{R}$ is the set of real numbers, is called negligible if for all ${c\in\mathbb{R}^+}$ there exist ${b_0\in\mathbb{N}}$ such that for all integers $b\ge b_0$ we have ${|\nu(b)|<\tfrac{1}{b^c}}$ \cite[Sect.~2.3.1]{Boneh2020AppliedCryptography}. In other words, $\nu(b)$ is negligible if it approaches zero faster than any inverse polynomial.
\end{definition}
Next, we define the semantic security notion.
\begin{definition}[Semantic security \cite{Probabilistic1984Goldwasser}]\label{semantic-def}
Consider a message space comprising any two equiprobable messages $\bold{M}^{(0)}, \bold{M}^{(1)} \in \{0,1\}^m$. Message $\bold{M}^{(i)},~ i\in \{0,1\}$, is chosen with probability $\frac{1}{2}$, encrypted using a cryptographic scheme, and sent from the transmitter to the receiver. Let $\mathbb{DE}: \{0,1\}^m \rightarrow \{0,1\}$ denote the decoding algorithm of an adversary with ``bounded'' computational power, and let $\kappa$ represent the security parameter.
The cryptographic scheme provides semantic security if for any decoding algorithm $\mathbb{DE}$, 
the probability that the adversary can compute bit $i$ from the ciphertext is better than a random guess by at most a negligible function in the security parameter $\kappa$.
In other words,  for any decoding algorithm $\mathbb{DE}$, the advantage of the adversary in distinguishing between the encryption of the two messages is smaller than a negligible function in the security parameter $\kappa$, denoted as $\nu(\kappa)$, i.e., 
\begin{align}\label{Def_s_seman}
&\bigg|\mathrm{Pr}\left(\mathbb{DE}[\bold{C}^{(1)}]=1\right)-\mathrm{Pr}\left(\mathbb{DE}[\bold{C}^{(0)}]=1\right)\bigg| < \nu(\kappa),
\end{align}
where $\bold{C}^{(i)}$ is the ciphertext corresponding to the message $\bold{M}^{(i)}$.
\end{definition}

   
%
   %

\begin{remark}
 It is necessary to set the security parameter $\kappa$ large enough, according to the system parameters, to achieve an appropriate bound (i.e., a small enough value) on the adversary's advantage.  
\end{remark}

\begin{definition}[{Unconditional security}]

If a cryptographic scheme ensures indistinguishability of encryptions of any two known messages against an adversary with ``unbounded'' computational power, i.e., in Definition~\ref{semantic-def}, the adversary is allowed to have unbounded computational power; hence, providing the absolute version of semantic security \cite{AumannZongRabin02, DingRabin02}, it is termed unconditionally secure 
\cite{AumannZongRabin02, AumannRabinAR, DziembowskiMaurer02, DingRabin02}, \cite[Sect.~2.2]{8584398}.
%
\end{definition}
\begin{remark}
Following the literature \cite[Sect.~2.2]{8584398}, \cite{AumannZongRabin02, AumannRabinAR, DziembowskiMaurer02, DingRabin02}, we differentiate between perfect security and unconditional security. Perfect security, provided by the one-time pad scheme \cite{Shannon}, ensures that an adversary with unbounded computational power gains no information about the message. However, it is not practical. Unconditional security ensures that, regardless of the computational power of the adversary and advances in computing technology and code-breaking algorithms, the adversary cannot gain even one bit of information about the message, except for an exponentially small probability in a security parameter. Thus, unconditionally secure schemes ensure everlasting security \cite{Maurer, AumannZongRabin02, AumannRabinAR, DziembowskiMaurer02, DingRabin02}.
\end{remark}

Next, we present a summary of the main results. First, we present the security results for a single message transmission, then, we show how to generalize the scheme for multiple message transmissions and provide the corresponding security results.

\subsection{Summary of the Main Results}\label{R The Multi-user IoBT System}
In the following theorem, we show that the provided encryption scheme is absolutely semantically secure. In other words, Theorem~\ref{SemanCrol_S} states that via using the provided scheme, for any pair of devices in the system, an all-powerful adversary with unbounded computation power cannot get even one bit of information about the transmitted message between any pair of devices, except with an \textit{exponentially} small probability in the security parameter $k$. 

\begin{theorem}
    \label{SemanCrol_S} 
For any $(q,l)\in\mathcal{Q}$, consider a message space consisting of any two equiprobable messages $\bold{M}_{(q,l)}^{(0)},~\bold{M}_{(q,l)}^{(1)}$ of lengths $m$ with $k<m<\lfloor\dfrac{n+1}{2}\rfloor$. 
Let $\mathbb{DE}: \{0,1\}^m\rightarrow\{0,1\}$ denote the decoding algorithm of an adversary with unbounded computation power. If the final key $\bold{X}_{(q,l)}$ is used for encryption, following the encryption scheme presented in Algorithm~\ref{Alg_Prot}, the advantage of the adversary in distinguishing  between the encryption of the two messages is upper-bounded as
\begin{align}\nonumber
&\bigg|\mathrm{Pr}\left(\mathbb{DE}\left[\bold{M}_{(q,l)}^{(1)}\oplus\bold{X}_{(q,l)}\right]=1\right)-\\&\nonumber\mathrm{Pr}\left(\mathbb{DE}\left[\bold{M}_{(q,l)}^{(0)}\oplus\bold{X}_{(q,l)}\right]=1\right)\bigg|<
\\&2W\left(\dfrac{(2m-1)^k}{n^k}+\dfrac{1}{2^{m}}\right),
\end{align}
{where $W=(U^2-U)/2$.
}
\end{theorem} 
\begin{proof}
See Section~\ref{Proof_SemanCrol_S}.
\end{proof}

Next, 
we show that the scheme can be used to securely exchange ${\eta_{\text{max}}}$ messages with length $m$ between any pair of devices $(q,l)$ via using the same binary random matrix $\boldsymbol\alpha$ and the pair-wise secret key $\bold{Z}_{(q,l)}$ given that $\eta_{\text{max}}m\le{\lfloor\frac{n+1}{2}\rfloor}$. More specifically, given that $\eta \le\eta_{\text{max}}$, to exchange the $\eta$th message with length $m$ between devices $q$ and $l$, they compute the final key $\bold{X}^{\eta}_{(q,l)}=(X_{(q,l),1}^{\eta},\ldots,X_{(q,l),m}^{\eta})$, where the $i$th bit of the final key $\bold{X}^{\eta}_{(q,l)}$, $X_{(q,l),i}^{\eta}$,  is computed as
\begin{align}\label{eq2ql2}
X_{(q,l),i}^{\eta}
=\bigoplus_{j=1}^k\boldsymbol{\alpha}[j,Z_{(q,l),j}+_n\left(m(\eta-1)+{(i-1)}\right)].
\end{align}
%
%
In Theorem~\ref{SemanCrol}, we show that the scheme can be used to securely exchange multiple messages between any pair of devices.

\begin{theorem}
    \label{SemanCrol} 
For any $(q,l)\in\mathcal{Q}$, consider a message space consisting of any two equiprobable messages $\bold{M}_{(q,l)}^{(0)},~\bold{M}_{(q,l)}^{(1)}$ of lengths $m$ with $k<m$. If the final key $\bold{X}_{(q,l)}^{{\eta}}$, where $1\le{\eta}\le\eta_{\text{max}}$ with $\eta_{\text{max}}m\le{\lfloor\frac{n+1}{2}\rfloor}$, is used for encryption, 
 the advantage of an adversary with unbounded computation power in distinguishing  between the encryption of the two messages is upper-bounded as
\begin{align}\nonumber
&\bigg|\mathrm{Pr}\left(\mathbb{DE}\left[\bold{M}_{(q,l)}^{(1)}\oplus\bold{X}_{(q,l)}^{{\eta}}\right]=1\right)-\\&\nonumber\mathrm{Pr}\left(\mathbb{DE}\left[\bold{M}_{(q,l)}^{(0)}\oplus\bold{X}_{(q,l)}^{{\eta}}\right]=1\right)\bigg|<
\\&2W\left(\dfrac{(2\eta_{\text{max}}m-1)^k}{n^k}+\dfrac{\eta_{\text{max}}}{2^{m}}\right).
\end{align}
\end{theorem} 
\begin{proof}
See Section~\ref{Proof_SemanCrol_S}.
\end{proof}

\begin{remark}
\label{newcroll3}
It is worth noting that if a pair of devices wants to exchange a larger amount of data during the mission, they could establish more than one pair-wise secret key and use each one separately to exchange $\eta_{\text{max}}$ messages. Note that each key needs to be chosen uniformly at random and independently of other keys.  In this situation, establishing one more pair-wise secret key is analogous to having one more pair of devices in the system that have established the pair-wise secret key to communicate. Thus, if  $\lambda_{(q,l)}$ denotes the number of pair-wise secret keys established between pair $(q,l)$, the results  in Theorems~\ref{SemanCrol_S} and \ref{SemanCrol} 
are updated by replacing $W$ by $\sum_{\forall (q,l)\in \mathcal{Q}}\lambda_{(q,l)}$.
%
%
\end{remark}


Next, we define the secrecy gain of the scheme, measuring the number of bits that a device can securely exchange with other devices by using a certain amount of stored secret bits.
\begin{definition}\label{def01}
Let $\Gamma_q$ denote the secrecy gain of the provided scheme for device $q$, defined as 
the ratio of the number of bits that device $q$ can securely exchange with other devices and the number of stored secret bits required on the device.
\end{definition}
    Using Definition~\ref{def01}, the secrecy gain $\Gamma_q$ is given as
\begin{align}\label{secgain}
   \Gamma_q&\stackrel{(a)}{=}\dfrac{\sum_{l\in\mathcal{U}\setminus \{q\}}\lambda_{(q,l)}m\eta_{\text{max}}}{\sum_{l\in\mathcal{U}\setminus \{q\}}\lambda_{(q,l)}k\log n+kn},
\end{align}
where the numerator of the right-hand side of $(a)$ comes from Theorem~\ref{SemanCrol} and Remark~\ref{newcroll3}, i.e., for each pair-wise secret key established between pair $(q,l)\in \mathcal{Q}$, $\eta_{\text{max}}$ messages with length $m$ can be securely exchanged, and the denominator of the right-hand side of $(a)$ comes from the fact that each pair-wise key has length $k\log n$ bits and the size of the shared binary random matrix $\boldsymbol{\alpha}$ is $kn$ bits. 


Similar to the secrecy gain $\Gamma_q$ defined for device $q$ in Definition~\ref{def01}, the \textit{system} secrecy gain for the considered IoBT system can be defined as the ratio of the total number of bits that can be securely exchanged across all device pairs to the number of secret bits used in the system, i.e., the random matrix $\boldsymbol{\alpha}$ and the pair-wise random keys. Let $\Gamma$ denote the system secrecy gain; then, it is characterized as follows:
\begin{align}\label{syssecgain}
   \Gamma&\stackrel{}{=}\dfrac{\sum_{(q,l)\in\mathcal{Q}}\lambda_{(q,l)}m\eta_{\text{max}}}{\sum_{(q,l)\in\mathcal{Q}}\lambda_{(q,l)}k\log n+kn}.
\end{align}

 \begin{example}\label{ex1}
 Consider an IoBT system with the following parameters  $n=2^{33}$ (which is around $1.075$ GB), length of each message $m=2^{10}$,  security parameter $k=46$, number of devices in the system $U=256$, maximum number of messages $\eta_{\text{max}}=2^{20}$, and number of established pair-wise secret keys between each pair  $\lambda_{(q,l)}=128,$ for all $(q,l)\in \mathcal{Q}$. 
 Then, the scheme enables all the possible pairs of devices in the system to exchange  around $17.1799$ GB, and the  advantage of an all-powerful adversary in distinguishing between the encryption of any two equiprobable messages is less than $2^{-69}$
 (see Theorem~\ref{SemanCrol}).
 The secrecy gain of each device is around $\Gamma_q\approx88.6845$, and the system secrecy gain is  $\Gamma=1.1174\times 10^4$.
\end{example}

Now, we relax the assumption that the binary random matrix $\boldsymbol\alpha$ is securely shared among the devices, i.e., we assume that the adversary gets access to the random matrix.
In addition, we allow the adversary to conduct a chosen plaintext attack (CPA) or chosen ciphertext attack (CCA) on the secret key $\bold{Z}_{(q,l)}$. More specifically, the adversary is allowed to have a plaintext--ciphertext pair $(\bold{M}_a,\bold{C}_a)$, where $\bold{C}_a$ is the encryption of the message $\bold{M}_a$ using the binary random matrix $\boldsymbol\alpha$ and secret key $\bold{Z}_{(q,l)}$. Thus, the adversary is provided with the final key $\bold{X}_a$, where $\bold{X}_a=\bold{C}_a\oplus\bold{M}_a$, and the whole bits of the binary random matrix $\boldsymbol\alpha$. Next, using the same binary random matrix $\boldsymbol\alpha$ and the same secret key $\bold{Z}_{(q,l)}$, message $\bold{M}_{(q,l)}$ is encrypted as $\bold{C}_{(q,l)}=\bold{M}_{(q,l)}\oplus\bold{X}_{(q,l)}$ and $\bold{C}_{(q,l)}$ is given to the adversary. We are interested in characterizing the computational complexity of decrypting the ciphertext $\bold{C}_{(q,l)}$ by the adversary using the binary random matrix $\boldsymbol\alpha$, and $\bold{X}_a$. 
This is carried out in the following theorem.
\begin{theorem}\label{ch-pl-theor}
For any secret key $\bold{Z}_{(q,l)}{\leftarrow} \mathbb{Z}_n^k$, for any random binary matrix $\boldsymbol\alpha$ revealed to the adversary, and for any plaintext--ciphertext pair $(\bold{M}_a,\bold{C}_a)$, captured by a CPA or CCA,
    the adversary can decrypt the ciphertext $\bold{C}_{(q,l)}$, encrypted by the random binary matrix $\boldsymbol\alpha$ and secret key $\bold{Z}_{(q,l)}$, 
    with a probability greater than $1-\frac{1}{2^{m}}$ by testing on average $2^{k\log n -1}$ secret keys.  
\end{theorem}
\begin{proof}
    See Section~\ref{Security of the protocol with Public Random Stings}.
\end{proof}

For example, consider $n=2^{33}$ and 
 $k=46$, then, using Theorem~\ref{ch-pl-theor},  the adversary requires to test on average $2^{1517}$ keys to recover the secret key. Thus, the scheme is computationally secure against the key recovery attack. 

\section{Proof of  Security}\label{Multi-user Systems}
In this section, we present the proofs of Theorems~\ref{SemanCrol_S} and \ref{SemanCrol}.

\subsection{Proof of Theorem~\ref{SemanCrol_S}}\label{Proof_SemanCrol_S}
To prove Theorem~\ref{SemanCrol_S}, we first provide the following lemma showing the relation between the advantage of the adversary and the probability of having a one-time pad to encrypt a message, i.e., { $\bold{X}_{(q,l)}$  is uniformly distributed, independent of any other final keys, and it is used once in the system.}
\begin{lemma}\label{lemnewp3}
  Let $\mathcal{Y}$ denote the event where the adversary can distinguish between the encryption of two equiprobable messages $\bold{M}_{(q,l)}^{(0)}$ and $\bold{M}_{(q,l)}^{(1)}$, then, the advantage of the adversary in distinguishing between the encryption of the two messages is characterized as
 \begin{align}\nonumber
&\mathrm{Pr}(\mathcal{Y})=\dfrac{1}{2}\bigg(1+\mathrm{Pr}\left(\mathbb{DE}\left[\bold{M}_{(q,l)}^{(1)}\oplus\bold{X}_{(q,l)}\right]=1\right)-\\& ~~~~~~~~~~~~~~~~~~~~~~\mathrm{Pr}\left(\mathbb{DE}\left[\bold{M}_{(q,l)}^{(0)}\oplus\bold{X}_{(q,l)}\right]=1\right)\bigg). 
  \end{align}
\end{lemma}
\begin{proof}
By using the law of total probability, the probability of the event $\mathcal{Y}$ is characterized as
\begin{align}\nonumber
    \mathrm{Pr}(\mathcal{Y})&=\mathrm{Pr}(\mathcal{Y}|~ \text{message}~\bold{M}_{(q,l)}^{(1)}~ \text{was transmitted})\times\\&\nonumber~~~~~~~~~~~~~~~~~~~~~\mathrm{Pr}(\text{message}~\bold{M}_{(q,l)}^{(1)}~ \text{was transmitted})\\&\nonumber+\mathrm{Pr}(\mathcal{Y}|~ \text{message}~\bold{M}_{(q,l)}^{(0)}~ \text{was transmitted})\times\\&\nonumber~~~~~~~~~~~~~~~~~~~~~\mathrm{Pr}(\text{message}~\bold{M}_{(q,l)}^{(0)}~ \text{was transmitted})\\&\nonumber\stackrel{(a)}{=}\dfrac{1}{2}\mathrm{Pr}\left(\mathbb{DE}\left[\bold{M}_{(q,l)}^{(1)}\oplus\bold{X}_{(q,l)}\right]=1\right)+\\&\nonumber~~~~~~~~~~~~~~~~~~~~~\dfrac{1}{2}\mathrm{Pr}\left(\mathbb{DE}\left[\bold{M}_{(q,l)}^{(0)}\oplus\bold{X}_{(q,l)}\right]=0\right)\\&\nonumber\stackrel{(b)}{=}\dfrac{1}{2}\bigg(1+\mathrm{Pr}\left(\mathbb{DE}\left[\bold{M}_{(q,l)}^{(1)}\oplus\bold{X}_{(q,l)}\right]=1\right)-\\&\nonumber~~~~~~~~~~~~~~~~~~~~~\mathrm{Pr}\left(\mathbb{DE}\left[\bold{M}_{(q,l)}^{(0)}\oplus\bold{X}_{(q,l)}\right]=1\right)\bigg),
\end{align}
where $(a)$ follows because i)  each message is chosen with probability $\frac{1}{2}$, and ii) probability of distinguishing between the encryption of the two messages when message $\bold{M}_{(q,l)}^{(i)},~i\in\{0,1\},$ was transmitted is equal to $\mathrm{Pr}\left(\mathbb{DE}\left[\bold{M}_{(q,l)}^{(i)}\oplus\bold{X}_{(q,l)}\right]=i\right)$, and $(b)$ comes from substituting $\mathrm{Pr}\left(\mathbb{DE}\left[\bold{M}_{(q,l)}^{(0)}\oplus\bold{X}_{(q,l)}\right]=0\right)=1-\mathrm{Pr}\left(\mathbb{DE}\left[\bold{M}_{(q,l)}^{(0)}\oplus\bold{X}_{(q,l)}\right]=1\right)$ into $(a)$.
\end{proof}
%
{
%
Let $\bar{\mathcal{C}}_{(q,l)}$ denote the event where the final key $\bold{X}_{(q,l)}$ computed by  Algorithm~\ref{Alg_Prot} is a one-time pad.
} 
Then, by using  Lemma~\ref{lemnewp3}, an upper-bound on the advantage of the adversary in distinguishing between the encryption of the two messages is given as
\begin{align}\nonumber \bigg|&\mathrm{Pr}\left(\mathbb{DE}\left[\bold{M}_{(q,l)}^{(1)}\oplus\bold{X}_{(q,l)}\right]=1\right)-\\&\nonumber\mathrm{Pr}\left(\mathbb{DE}\left[\bold{M}_{(q,l)}^{(0)}\oplus\bold{X}_{(q,l)}\right]=1\right)\bigg|\stackrel{(a)}{=}\big|2\mathrm{Pr}(\mathcal{Y})-1\big|\\&\nonumber
\stackrel{}{=}\!\bigg|2\left(\mathrm{Pr}(\mathcal{Y}|\mathcal{C}_{(q,l)})\mathrm{Pr}(\mathcal{C}_{(q,l)})\!+\!\mathrm{Pr}(\mathcal{Y}|\bar{\mathcal{C}}_{(q,l)})\mathrm{Pr}(\bar{\mathcal{C}}_{(q,l)})\right)\!-\!1\bigg|.
\\&\label{Sum_indi_prob0}
\end{align}
where $(a)$ follows from Lemma~\ref{lemnewp3}.
Now, we consider two cases: i) $\mathrm{Pr}(\mathcal{Y})\ge\dfrac{1}{2}$ and ii) $\mathrm{Pr}(\mathcal{Y})<\dfrac{1}{2}$. For the case $\mathrm{Pr}(\mathcal{Y})\ge\dfrac{1}{2}$, we have 
\begin{align}\nonumber 
&\big|2\mathrm{Pr}(\mathcal{Y})-1\big|=2\mathrm{Pr}(\mathcal{Y})-1\\&\nonumber
2\left(\mathrm{Pr}(\mathcal{Y}|\mathcal{C}_{(q,l)})\mathrm{Pr}(\mathcal{C}_{(q,l)})\!+\!\mathrm{Pr}(\mathcal{Y}|\bar{\mathcal{C}}_{(q,l)})\mathrm{Pr}(\bar{\mathcal{C}}_{(q,l)})\right)\!-\!1
\\&\nonumber
\stackrel{}{\le}2\left(\mathrm{Pr}(\mathcal{C}_{(q,l)})+\mathrm{Pr}(\mathcal{Y}|\bar{\mathcal{C}}_{(q,l)})\right)-1
\\&\label{Sum_indi_prob00}
\stackrel{(a)}{=}2\mathrm{Pr}(\mathcal{C}_{(q,l)}),
\end{align}
where 
$(a)$ follows because when the final key  $\bold{X}_{(q,l)}$ is a one-time pad, i.e., under the event $\bar{\mathcal{C}}_{(q,l)}$, the probability of distinguishing between the encryption of the two messages is $\frac{1}{2}$, i.e., $\mathrm{Pr}(\mathcal{Y}|\bar{\mathcal{C}}_{(q,l)})=\frac{1}{2}$.
%
For the case $\mathrm{Pr}(\mathcal{Y})<\dfrac{1}{2}$, we have 
\begin{align}\nonumber 
&\big|2\mathrm{Pr}(\mathcal{Y})-1\big|=1-2\mathrm{Pr}(\mathcal{Y})\\&\nonumber
=1\!-\!2\left(\mathrm{Pr}(\mathcal{Y}|\mathcal{C}_{(q,l)})\mathrm{Pr}(\mathcal{C}_{(q,l)})\!+\!\mathrm{Pr}(\mathcal{Y}|\bar{\mathcal{C}}_{(q,l)})\mathrm{Pr}(\bar{\mathcal{C}}_{(q,l)})\right)
\\&\nonumber
=1\!-\!2\left(\mathrm{Pr}(\mathcal{Y}|\mathcal{C}_{(q,l)})\mathrm{Pr}(\mathcal{C}_{(q,l)})\!+\dfrac{1}{2}(1-\mathrm{Pr}({\mathcal{C}}_{(q,l)}))\right)
\\&\nonumber
\stackrel{(a)}{=}\mathrm{Pr}(\mathcal{C}_{(q,l)})\left(1-2\mathrm{Pr}(\mathcal{Y}|{\mathcal{C}}_{(q,l)})\right)
\\&\label{Sum_indi_prob000}
\stackrel{}{<}\mathrm{Pr}(\mathcal{C}_{(q,l)}),
\end{align}
where $(a)$ follows because i) $\mathrm{Pr}(\mathcal{Y}|\bar{\mathcal{C}}_{(q,l)})=\frac{1}{2}$, and ii) $\mathrm{Pr}(\bar{\mathcal{C}}_{(q,l)})=1-\mathrm{Pr}({\mathcal{C}}_{(q,l)})$.

From \eqref{Sum_indi_prob00} and \eqref{Sum_indi_prob000}, we have 
\begin{align}\nonumber \bigg|&\mathrm{Pr}\left(\mathbb{DE}\left[\bold{M}_{(q,l)}^{(1)}\oplus\bold{X}_{(q,l)}\right]=1\right)-\\&\label{Sum_indi_prob}
\mathrm{Pr}\left(\mathbb{DE}\left[\bold{M}_{(q,l)}^{(0)}\oplus\bold{X}_{(q,l)}\right]=1\right)\bigg|<2\mathrm{Pr}(\mathcal{C}_{(q,l)}).
\end{align}
Thus, our remaining task is to characterize  $\mathrm{Pr}(\mathcal{C}_{(q,l)})$, which is carried out in the following proposition.

\begin{proposition}\label{One-time-Theorem}
For any binary random matrix $\boldsymbol\alpha{\leftarrow}\{0,1\}^{k\times n}$, for any pair of devices $(q, l) \in\mathcal{Q}$, the final secret key $\bold{X}_{(q,l)}\in\{0,1\}^m$ derived by the proposed protocol in Algorithm~\ref{Alg_Prot} is a one-time pad, except for an exponentially small probability in the security parameter $k$ 
where $k<m<\lfloor\dfrac{n+1}{2}\rfloor$.
More specifically, for any pair of devices $(q,l)$, an upper-bound on $\mathrm{Pr}(\mathcal{C}_{(q,l)})$ is given as
\begin{align}\label{theo-upper-exp}
    \mathrm{Pr}(\mathcal{C}_{(q,l)})<W\left(\dfrac{(2m-1)^k}{n^k}+\dfrac{1}{2^m}\right).
\end{align}
 \end{proposition}
 \begin{proof}
     First, we introduce some necessary notations for the proof in the following definition. 
\begin{definition}\label{Def1}
Let $\bold{S}_{(q,l)}^{(i)}=
\bold{Z}_{(q,l)}+_n(i-1)\bm{1}$, where $\bm{1}=(1,\ldots,1)\in \mathbb{Z}_n^k$, denote the sub-key that is used to compute the $i$th bit of the final key $\bold{X}_{(q,l)}$. Recall from \eqref{eq2ql} that components of $\bold{S}_{(q,l)}^{(i)}$, i.e., $\left(Z_{(q,l),1}+_n(i-1),\ldots,Z_{(q,l),k}+_n(i-1)\right)$, determine the location of bits in $\boldsymbol{\alpha}$ that are used to compute the $i$th bit of the final key $\bold{X}_{(q,l)}$. Let $\mathcal{S}_{(q,l)}$ denote the $m$-tuple containing the $m$ sub-keys used to compute the final key $\bold{X}_{(q,l)}$, i.e., ${\mathcal{S}_{(q,l)}=\left(\bold{S}_{(q,l)}^{(1)},\ldots,\bold{S}_{(q,l)}^{(m)}\right)}$. 
\end{definition}

Next, we show that: i)
all the tuples of sub-keys, $\mathcal{S}_{(1,2)},\mathcal{S}_{(1,3)},\ldots,\mathcal{S}_{({U-1},{U})},$ 
are statistically independent and each sub-key is chosen uniformly at random from $\mathbb{Z}_n^k$, which is shown in Lemma~\ref{pair independence}, 
ii) for any pair of devices $(q,l)$, components of $\bold{X}_{(q,l)}\in\{0,1\}^m$, i.e., $(X_{(q,l),1},\ldots,X_{(q,l),m})$ are independent of each other, which is shown in Lemma~\ref{Component independence}, and iii) for any pair of devices $(q,l)$, components of $\bold{X}_{(q,l)}\in\{0,1\}^m$ are equally likely to take on $0$ or $1$  as their value, which is shown in Lemma~\ref{Component equallylikly}. 
Finally, we show that the final key $\bold{X}_{(q,l)}$  is independent of any other final keys and is used once in the system, i.e., it is a one-time pad, except for an exponentially small probability in the security parameter $k$.

\begin{lemma}\label{pair independence}
Each sub-key $\bold{S}_{(q,l)}^{(i)}$ for  ${i=1,\ldots, m}$ and $(q,l)\in\mathcal{Q}$, is chosen uniformly at random from $\mathbb{Z}_n^k$, i.e., $\bold{S}_{(q,l)}^{(i)}{\leftarrow} \mathbb{Z}_n^k$, and all the tuples of sub-keys, $\mathcal{S}_{(1,2)},\mathcal{S}_{(1,3)},\ldots,\mathcal{S}_{({U-1},{U})},$ 
are statistically independent. 
\end{lemma}
\begin{proof}
Recall that for each pair of devices $(q,l)$, the $i$th sub-key, which is the $i$th component of $\mathcal{S}_{(q,l)}$, is computed as $\bold{S}_{(q,l)}^{(i)}=\bold{Z}_{(q,l)}+_n(i-1)\bm{1}$. Since  $\bold{Z}_{(q,l)}{\leftarrow} \mathbb{Z}_n^k$, 
we have $\bold{S}_{(q,l)}^{(i)}{\leftarrow} \mathbb{Z}_n^k$ for  ${i=1,\ldots, m}$.
Thus, for each pair $(q,l)$, the components of the tuple of  sub-keys ${\mathcal{S}_{(q,l)}=\left(\bold{S}_{(q,l)}^{(1)},\ldots,\bold{S}_{(q,l)}^{(m)}\right)}$ are chosen uniformly at random from $\mathbb{Z}_n^k$ and they depend only on the secret key $\bold{Z}_{(q,l)}$. Since each set of sub-keys depends only on the corresponding pair-wise secret key and all the pair-wise secret keys are statistically independent, 
%
%
 all the tuples of sub-keys, $\mathcal{S}_{(1,2)},\mathcal{S}_{(1,3)},\ldots,\mathcal{S}_{({U-1},{U})},$ 
are statistically independent.
\end{proof}

\begin{lemma}\label{Component independence}
    For any pair of devices $(q,l)\in\mathcal{Q}$, for any $\bold{Z}_{(q,l)}{\leftarrow} \mathbb{Z}_n^k$, components of $\bold{X}_{(q,l)}\in\{0,1\}^m$, i.e., $(X_{(q,l),1},\ldots,X_{(q,l),m})$, are statistically independent.
\end{lemma}
\begin{proof} 
Recall that $X_{(q,l),i}=\bigoplus_{j=1}^k\boldsymbol{\alpha}[j,Z_{(q,l),j}+_n(i-1)]$.
We show that for any two distinct $i$ and $i'$, the sets of bits of binary random matrix $\boldsymbol\alpha$ that are used to compute  $X_{(q,l),i}$ and $X_{(q,l),i'}$ have no bit in common, thus,  the components of $\bold{X}_{(q,l)}$ are computed by using different bits of the binary random matrix $\boldsymbol\alpha$ and consequently, they are statistically independent.
By looking at the procedure of computing $X_{(q,l),i}$ and $X_{(q,l),i'}$, we can see that the sets of bits of the binary random matrix $\boldsymbol\alpha$ that are used to compute  $X_{(q,l),i}$ and $X_{(q,l),i'}$ have no bit in common if the components of the $k$-tuples $\bold{D}_{ii'}=\bold{S}_{(q,l)}^{(i)}-\bold{S}_{(q,l)}^{(i')}$ are all non-zero. Using Definition~\ref{Def1},  we have $\bold{D}_{ii'}=(i-i')\bm{1}$, which is a $k$-tuples with non-zero components for any two distinct $i$ and $i'$.
\end{proof}

\begin{lemma}\label{Component equallylikly}
    For any pair of devices $(q,l)\in\mathcal{Q}$, components of ${\bold{X}_{(q,l)}\in\{0,1\}^m}$ are equally likely to take on $0$ or $1$ as their value.
\end{lemma}
\begin{proof}
Let $p$ and $p'$ denote the probability of $1$ and $0$ in the binary random matrix $\boldsymbol\alpha$, respectively. Then, since $\bold{S}_{(q,l)}^{(i)}$'s are uniform over $\mathbb{Z}_{n}^{k}$, the probability that the $i$th component of $\bold{X}_{(q,l)}$, i.e., $X_{(q,l),i}$, for $i=1,\ldots,m$, takes on  $1$ or $0$ as its value, are 
calculated as follows:
 \begin{align}\nonumber
&\mathrm{Pr}(X_{(q,l),i}=1)=\sum_{j~ \mathrm{odd}}{k \choose j}p^j{p'}^{k-j},\\&
\mathrm{Pr}(X_{(q,l),i}=0)=\sum_{j~ \mathrm{even}}{k \choose j}p^j{p'}^{k-j}.
\end{align}
    Using the Binomial theorem \cite[Page~162]{BGraham1994Concrete}, we have 
    \begin{align}\nonumber
        &({p'}+p)^k=\sum_{j~ \mathrm{even}}{k \choose j}p^j {p'}^{k-j}+\sum_{j~ \mathrm{odd}}{k \choose j}p^j {p'}^{k-j},\\&
        ({p'}-p)^k=\sum_{j~ \mathrm{even}}{k \choose j}p^j {p'}^{k-j}-\sum_{j~\mathrm{odd}}{k \choose j}p^j {p'}^{k-j}.
    \end{align}
    By applying mathematical manipulations and using the fact that $p+{p'}=1$, we have 
    \begin{align}\nonumber
        \mathrm{Pr}(X_{(q,l),i}=1)&=\sum_{j~ \mathrm{odd}}{k \choose j}p^j{p'}^{k-j}\\&\nonumber=\dfrac{1-({p'}-p)^k}{2},\\&\nonumber
        \hspace{-2.4cm}\mathrm{Pr}(X_{(q,l),i}=0)=\sum_{j~ \mathrm{even}}{k \choose j}p^j{p'}^{k-j}\\&=\dfrac{1+({p'}-p)^k}{2}.
    \end{align}
    Using the  assumption that we have 
    ${p={p'}=\frac{1}{2}}$, the probabilities are given as ${\mathrm{Pr}(X_{(q,l),i}=1)=\mathrm{Pr}(X_{(q,l),i}=0)=\frac{1}{2}}$ for $i=1,\ldots,m$.
\end{proof}

 
 The final key $\bold{X}_{(q,l)}$ for  $(q,l) \in\mathcal{Q}$, is a one-time pad if it is  i) chosen uniformly at random from $\{0,1\}^m$, 
 ii) independent of all other final keys $\bold{X}_{(q',l')}$ for $(q',l')\in\mathcal{Q}$ with $(q,l)\ne(q',l')$, 
 and iii) used only once in the whole system.
 %
%
%
In Lemmas~\ref{Component independence} and \ref{Component equallylikly}, we showed that using the provided scheme in Algorithm~\ref{Alg_Prot}, any final key $\bold{X}_{(q,l)}$, is a uniformly distributed random variable, i.e., ${\bold{X}_{(q,l)}{\leftarrow} \{0,1\}^m}$ for  $(q,l) \in\mathcal{Q}$. 
We show that $\mathrm{Pr}(\mathcal{C}_{(q,l)})$, i.e., the probability that the final key $\bold{X}_{(q,l)}$ is not a one-time pad, is exponentially small in the security parameter $k$. To this end, first, we present the following lemma known as the Crypto lemma \cite{https://doi.org/10.48550/arxiv.cs/0409053}. 
\begin{lemma}\label{cryptolem}
Let $\mathcal{A}$ be a compact abelian group with group operation $\pmb{+}$, and $B$ and $C$ be random variables over $\mathcal{A}$ where $B$ is independent of $C$ and uniform over $\mathcal{A}$. Then, $E = B \pmb{+} C$ is independent of $C$ and uniform over $\mathcal{A}$.  
\end{lemma}
\begin{proof}
See \cite[Lemma~2]{https://doi.org/10.48550/arxiv.cs/0409053}.
\end{proof}
According to Lemma~\ref{cryptolem}, 
we can see that $\bold{X}_{(q,l)}$ is independent of all other final keys 
if to compute each bit of the final key $\bold{X}_{(q,l)}$, we use at least one bit of the random binary matrix $\boldsymbol{\alpha}$ that is not used to compute any other final keys. In other words, each element of the 
tuple of sub-keys $\mathcal{S}_{(q,l)}=\left(\bold{S}_{(q,l)}^{(1)},\ldots,\bold{S}_{(q,l)}^{(m)}\right)$ selects bits from the random binary matrix $\boldsymbol{\alpha}$ in such a way that at least one of the bits is not selected by elements of any other 
tuple of sub-keys $\mathcal{S}_{(q',l')}=\left(\bold{S}_{(q',l')}^{(1)},\ldots,\bold{S}_{(q',l')}^{(m)}\right)$ for $(q',l')\in\mathcal{Q}$ with $(q,l)\ne(q',l')$.

We characterize  $\mathrm{Pr}(\mathcal{C}_{(q,l)})$ by means of the event $\mathcal{E}_{(q,l)}$ denoting the event where for \textit{at least} one bit of the final key $\bold{X}_{(q,l)}$, e.g., ${X}^i_{(q,l)}$, all the selected bits by the corresponding sub-key, 
i.e., $\bold{S}_{(q,l)}^{(i)}$, were used at least once in computing other final keys, and thus, $\bold{X}_{(q,l)}$ depends on other final keys. 
%
%
Then, according to the law of total probability, the probability that the final key $\bold{X}_{(q,l)}$ is not a one-time pad
is characterized as 
\begin{align}\nonumber
\mathrm{Pr}(\mathcal{C}_{(q,l)})&=\mathrm{Pr}(\mathcal{C}_{(q,l)}|\mathcal{E}_{(q,l)})\mathrm{Pr}(\mathcal{E}_{(q,l)})\\&\label{Pro-upper-I-0}+\mathrm{Pr}(\mathcal{C}_{(q,l)}|\mathcal{\bar{E}}_{(q,l)})\mathrm{Pr}(\mathcal{\bar{E}}_{(q,l)}),
\end{align}
where $\mathcal{\bar{E}}_{(q,l)}$ is the complementary event of $\mathcal{E}_{(q,l)}$. By ignoring the terms $\mathrm{Pr}(\mathcal{C}_{(q,l)}|\mathcal{E}_{(q,l)})$ and $\mathrm{Pr}(\mathcal{\bar{E}}_{(q,l)})$ in \eqref{Pro-upper-I-0}, an upper-bound on $\mathrm{Pr}(\mathcal{C}_{(q,l)})$ is given as
\begin{align}\label{Pro-upper-I0}
    \mathrm{Pr}(\mathcal{C}_{(q,l)})\le\mathrm{Pr}(\mathcal{E}_{(q,l)})+\mathrm{Pr}(\mathcal{C}_{(q,l)}|\mathcal{\bar{E}}_{(q,l)}).
\end{align}
Next, we characterize the values of probabilities $\mathrm{Pr}(\mathcal{E}_{(q,l)})$ and $\mathrm{Pr}(\mathcal{C}_{(q,l)}|\mathcal{\bar{E}}_{(q,l)})$. To this end, we consider a worst-case scenario in which all the possible pairs of devices in the system want to exchange one message, and thus, $\mathrm{Pr}(\mathcal{E}_{(q,l)})$ and $\mathrm{Pr}(\mathcal{C}_{(q,l)}|\mathcal{\bar{E}}_{(q,l)})$ have their highest values.

Characterizing the value of $\mathrm{Pr}(\mathcal{E}_{(q,l)})$: We start with characterizing the complementary event of $\mathcal{E}_{(q,l)}$, i.e., $\mathcal{\bar{E}}_{(q,l)}$. Under the event $\mathcal{\bar{E}}_{(q,l)}$, to compute each bit of the final key $\bold{X}_{(q,l)}$, the protocol uses at least one bit of the binary random matrix $\boldsymbol{\alpha}$ that is not used in computing final keys of all other pairs of devices.
{An extreme case under which the event $\mathcal{\bar{E}}_{(q,l)}$ occurs is that}  
the pair-wise secret keys of all possible pairs of devices, i.e., $\bold{Z}_{(q',l')},~\forall(q',l')\in\mathcal{Q}$ with $(q,l)\ne(q',l'),$ do not take a value with the following form
\begin{align}\label{new_eq_r}   
\bold{Z}_{(q,l)}+(f_1,f_2,\ldots,f_k),
\end{align}
where $f_i\in\{-m+1,-m+2,\cdots,m-1\}$ for $i=1,\ldots,k$.  Let $\mathcal{\bar{E}}^{(q',l')}_{(q,l)}$ denote the event where the uniformly chosen random pair-wise secret key $\bold{Z}_{(q',l')}$ does not take a value as in \eqref{new_eq_r}. Then,  probability of the event $\mathcal{\bar{E}}^{(q',l')}_{(q,l)}$  is  given as
\begin{align}\label{Pro-upper-II0}
\mathrm{Pr}(\mathcal{\bar{E}}^{(q',l')}_{(q,l)})\stackrel{}{=}\dfrac{n^k-{(2m-1)}^k}{n^k}.
\end{align}
Finally, an upper-bound on the probability of the event $\mathcal{E}_{(q,l)}$ is given as 
\begin{align}\nonumber
\mathrm{Pr}(\mathcal{E}_{(q,l)})&=1-\mathrm{Pr}(\mathcal{\bar{E}}_{(q,l)})\\&\nonumber\stackrel{(a)}{\le}1-\prod_{\forall (q',l'),(q',l')\ne (q,l)}\mathrm{Pr}(\mathcal{\bar{E}}^{(q',l')}_{(q,l)})
\\&\nonumber\stackrel{(b)}{=}
1-\left(1-\dfrac{(2m-1)^k}{n^k}\right)^{\left(\dfrac{U^2-U}{2}\right)-1}\\&\label{Pro-upper-I00}\stackrel{(c)}{<}\dfrac{(U^2-U)(2m-1)^k}{2n^k},
\end{align}
where {the multiplication on the right-hand side of} $(a)$ follows because 
all the pair-wise secret keys $\bold{Z}_{(1,2)},\bold{Z}_{(1,3)},\ldots,\bold{Z}_{({U-1},{U})},$  are statistically independent,
$(b)$ follows from \eqref{Pro-upper-II0} and the fact that the total number of pairs in the system except the pair $(q,l)$ is ${U \choose 2}-1=\frac{U^2-U}{2}-1$; 
and $(c)$ follows from the Bernoulli's inequality, i.e., $(1-t)^r>1-tr,$ for $0\le t\le1$ and $ r>1$ \cite[Sect. 1.3.3]{brannan2006first}, and ignoring some negative terms.

Characterizing the value of $\mathrm{Pr}(\mathcal{C}_{(q,l)}|\mathcal{\bar{E}}_{(q,l)})$: 
An upper-bound on $\mathrm{Pr}(\mathcal{C}_{(q,l)}|\mathcal{\bar{E}}_{(q,l)})$ is given as
\begin{align}\nonumber
\mathrm{Pr}(\mathcal{C}_{(q,l)}|\mathcal{\bar{E}}_{(q,l)})
&\stackrel{(a)}{\le}\dfrac{(U^2-U)/2-1}{2^m}\\&\label{Pro-upper-III}
<\dfrac{U^2-U}{2^{m+1}},
\end{align}
where 
$(a)$ follows because i) under the event $\mathcal{\bar{E}}_{(q,l)}$, the final key $\bold{X}_{(q,l)}$ is independent of all other final keys $\bold{X}_{(q',l')}$ with $(q,l)\ne(q',l')$,  ii) as shown in Lemmas~\ref{Component independence} and \ref{Component equallylikly}, the components of ${\bold{X}_{(q,l)}}$ are statistically independent and they are equally likely to take on $0$ and $1$ as their value, and iii) the total number of pairs in the system that could have the same final key as $\bold{X}_{(q,l)}$ is at most $(\frac{U^2-U}{2})-1$. 

Substituting the derived bounds in \eqref{Pro-upper-I00} and \eqref{Pro-upper-III} into  \eqref{Pro-upper-I0} completes the proof of Proposition~\ref{One-time-Theorem}. 
%
%
%
 \end{proof}

 Finally, substituting the bound in \eqref{theo-upper-exp} into \eqref{Sum_indi_prob} completes the proof of Theorem~\ref{SemanCrol_S}.

 \subsection{Proof of  Theorem~\ref{SemanCrol}}\label{Proof_SemanCrol}

The concept of the proof of Theorem~\ref{Th-exp-pro-2} is the same as that of Theorem~\ref{One-time-Theorem}. Thus, here, we avoid presenting details that can be understood by simply generalizing the corresponding part in the proof of Theorem~\ref{One-time-Theorem}. 

Following the same steps applied to derive the bound in  \eqref{Sum_indi_prob}, one can show that
\begin{align}\nonumber &\bigg|\mathrm{Pr}\left(\mathbb{DE}\left[\bold{M}_{(q,l)}^{(1)}\oplus\bold{X}^\eta_{(q,l)}\right]=1\right)-\\&\label{multi-m-n-p}
\mathrm{Pr}\left(\mathbb{DE}\left[\bold{M}_{(q,l)}^{(0)}\oplus\bold{X}^\eta_{(q,l)}\right]=1\right)\bigg|\le2\mathrm{Pr}(\mathcal{C}^\eta_{(q,l)}),
\end{align}
where $\mathcal{C}^\eta_{(q,l)}$  denotes the event where the final key $\bold{X}^\eta_{(q,l)}$ is not a one-time pad.
In the following proposition, we provide an upper-bound on $\mathrm{Pr}(\mathcal{C}^\eta_{(q,l)})$.

\begin{proposition}\label{Th-exp-pro-2}
For any pair of devices $(q,l)\in \mathcal{Q}$, for $1\le\eta \le\eta_{\text{max}}$ with $\eta_{\text{max}}m\le{\lfloor\frac{n+1}{2}\rfloor}$, an upper-bound on the probability that the final secret key $\bold{X}_{(q,l)}^\eta\in\{0,1\}^m$ with $k<m$ derived by the protocol is not a one-time pad is given as
%
\begin{align}\label{maxrta}
\mathrm{Pr}(\mathcal{C}^\eta_{(q,l)})< W\left(\dfrac{(2\eta_{\text{max}}m-1)^k}{n^k}+\dfrac{\eta_{\text{max}}}{2^m}\right).
\end{align}
\end{proposition}
\begin{proof}   
In the following, we first introduce some necessary notations and then present the proof.

\begin{definition}
Let $\bold{S}_{(q,l)}^{\eta,(i)}=
\bold{Z}_{(q,l)}+_n(m(\eta-1)+_ni-1)\bm{1}$, where $\bm{1}=(1,\ldots,1)\in \mathbb{Z}_n^k$, denote the sub-key that is used to compute the $i$th bit of the final key $\bold{X}^{\eta}_{(q,l)}$, and    $\mathcal{S}^{\eta}_{(q,l)}$ denote the $m$-tuple containing the $m$ sub-keys used to compute the final key $\bold{X}^{\eta}_{(q,l)}$, i.e., ${\mathcal{S}^{\eta}_{(q,l)}=\left(\bold{S}_{(q,l)}^{\eta,(1)},\ldots,\bold{S}_{(q,l)}^{\eta,(m)}\right)}$. 
\end{definition}

Following the same steps as for the proofs of Lemmas~\ref{Component independence} and \ref{Component equallylikly} and considering that $\eta\le \eta_{\text{max}}$, one can show that using the provided protocol in Algorithm~\ref{Alg_Prot}, the final key $\bold{X}^\eta_{(q,l)}$ is a uniformly distributed random variable, i.e., ${\bold{X}^\eta_{(q,l)}{\leftarrow} \{0,1\}^m}$ for $(q,l)\in\mathcal{Q}$. 
To derive an upper-bound on $\mathrm{Pr}(\mathcal{C}^\eta_{(q,l)})$, 
we consider a worst-case scenario where all the possible pairs of devices in the system want to exchange $\eta_{\text{max}}$ messages.

The final key $\bold{X}^\eta_{(q,l)}$ is not a one-time pad if either it depends on any other final key $\bold{X}^{\eta'}_{(q',l')}$ with ${((q,l),\eta)\ne((q',l'),\eta')}$,
for all $1\le\eta'\le\eta_{\text{max}}$ and $(q',l')\in \mathcal{Q}$ or it is independent of other final keys, but, it is used more than once in the whole system.
 Let $\mathcal{E}
^\eta_{(q,l)}$ denotes the event where for \textit{at least} one bit of the final key $\bold{X}^{\eta}_{(q,l)}$ all the selected bits by the corresponding sub-key
were used at least once in computing other final keys, and thus, $\bold{X}^{\eta}_{(q,l)}$  depends on other final keys.
Then, by using the law of total probability $\mathrm{Pr}(\mathcal{C}^\eta_{(q,l)})$ is characterized as
\begin{align}\nonumber
\mathrm{Pr}(\mathcal{C}^\eta_{(q,l)})&=\mathrm{Pr}(\mathcal{C}^\eta_{(q,l)}|\mathcal{E}^\eta_{(q,l)})\mathrm{Pr}(\mathcal{E}^\eta_{(q,l)})\\&\nonumber+\mathrm{Pr}(\mathcal{C}^\eta_{(q,l)}|\mathcal{\bar{E}}^\eta_{(q,l)})\mathrm{Pr}(\mathcal{\bar{E}}^\eta_{(q,l)}),\\&\label{th2Pro-upper-I-0}
\le\mathrm{Pr}(\mathcal{E}^\eta_{(q,l)})+\mathrm{Pr}(\mathcal{C}^\eta_{(q,l)}|\mathcal{\bar{E}}^\eta_{(q,l)}).
\end{align}
%
%
%
%
Next, we characterize the values of the probabilities $\mathrm{Pr}(\mathcal{E}^\eta_{(q,l)})$ and $\mathrm{Pr}(\mathcal{C}^\eta_{(q,l)}|\mathcal{\bar{E}}^\eta_{(q,l)})$. 

Characterizing the value of $\mathrm{Pr}(\mathcal{E}^\eta_{(q,l)})$: Since $\eta$ and $\eta'$ are less than or equal to $\eta_{\text{max}}$ and ${\eta_{\text{max}}m\le{\lfloor\frac{n+1}{2}\rfloor}}$,  $\mathcal{S}^{\eta}_{(q,l)}$ and $\mathcal{S}^{\eta'}_{(q,l)}$ for all $\eta'\ne\eta,$ choose different elements of the random binary matrix $\boldsymbol{\alpha}$ to compute $\bold{X}^{\eta}_{(q,l)}$ and $\bold{X}^{\eta'}_{(q,l)}$ for all $\eta'\ne\eta$, and thus, $\bold{X}^{\eta}_{(q,l)}$ and $\bold{X}^{\eta'}_{(q,l)}$ are independent. Thus, we need to investigate the dependency between $\bold{X}^{\eta}_{(q,l)}$ and the final keys of other pairs.
Similar to the proof of Theorem~1, we first characterize the event $\mathcal{\bar{E}}^\eta_{(q,l)}$, i.e., each bit of the final key $\bold{X}^{\eta}_{(q,l)}$ uses at least one bit of the binary random matrix $\boldsymbol{\alpha}$ that is not used for computing the $\eta_{\text{max}}$ final keys of all other pairs of devices.
{An extreme case under which the event $\mathcal{\bar{E}}^\eta_{(q,l)}$ occurs is that}
the pair-wise secret keys of all other pairs of devices, i.e., $\bold{Z}_{(q',l')},~\forall(q',l')\in\mathcal{Q}$ with $(q,l)\ne(q',l')$ do not take a value with the following form
\begin{align}\label{new_eq_r2}   
\bold{Z}_{(q,l)}+(\bar{f}_1,\bar{f}_2,\ldots,\bar{f}_k),
\end{align}
where $\bar{f}_i\in\{-m\eta_{\text{max}}+1,-m\eta_{\text{max}}+2,\cdots,m\eta_{\text{max}}-1\}$ for $i=1,\ldots,k$. Let $\mathcal{\bar{E}}^{(q',l'),T}_{(q,l)}$ denote the event where the pair-wise secret key $\bold{Z}_{(q',l')}$ does not take a value as in \eqref{new_eq_r2}. Then, probability of event $\mathcal{\bar{E}}^{(q',l'),T}_{(q,l)}$ is  given as
\begin{align}\label{Pro-upper-II00ii}
\mathrm{Pr}(\mathcal{\bar{E}}^{(q',l'),T}_{(q,l)})\stackrel{}{=}\dfrac{n^k-{(2\eta_{\text{max}}m-1)}^k}{n^k}.
\end{align}

%
Finally, using \eqref{Pro-upper-II00ii}, an upper-bound on the value of  $\mathrm{Pr}(\mathcal{E}^\eta_{(q,l)})$ is given as
\begin{align}\nonumber
\mathrm{Pr}(\mathcal{E}^\eta_{(q,l)})&=1-\mathrm{Pr}(\mathcal{\bar{E}}^\eta_{(q,l)})\\&\nonumber\stackrel{(a)}{\le}1-\prod_{\forall (q',l'),(q',l')\ne (q,l)}\mathrm{Pr}(\mathcal{\bar{E}}^{(q',l'),T}_{(q,l)})
\\&\nonumber\stackrel{}{=}1-\left(\dfrac{n^k-(2\eta_{\text{max}}m-1)^k}{n^k}\right)^{\left(\dfrac{U^2-U}{2}\right)-1}\\&\label{Pro-upper-I00I}\stackrel{}{<}\dfrac{(2\eta_{\text{max}}m-1)^k(U^2-U)}{2n^k},
\end{align}       
where $(a)$ follows because all the pair-wise secret keys are independent. 
%

Characterizing the value of $\mathrm{Pr}(\mathcal{C}^\eta_{(q,l)}|\mathcal{\bar{E}}^\eta_{(q,l)})$:
Under the event $\mathcal{\bar{E}}^\eta_{(q,l)}$, an upper-bound on $\mathrm{Pr}(\mathcal{C}^\eta_{(q,l)}|\mathcal{\bar{E}}^\eta_{(q,l)})$ is given as
\begin{align}\label{Pro-upper-IV-II-2}
\mathrm{Pr}(\mathcal{C}^{\eta}_{(q,l)}|\mathcal{\bar{E}}^\eta_{(q,l)})
\stackrel{(a)}{<}\dfrac{\eta_{\text{max}}(U^2-U)}{2^{m+1}},
\end{align}
where  $(a)$ follows because i) under the event $\mathcal{\bar{E}}^\eta_{(q,l)}$, the secret key $\bold{X}^{\eta}_{(q,l)}$ is independent of any other  $\bold{X}^{\eta'}_{(q',l')}$ with $((q,l),\eta)\ne((q',l'),\eta')$,  ii) the components of ${\bold{X}^{\eta}_{(q,l)}\in\{0,1\}^m}$ are statistically independent and they are equally likely to take on $0$ and $1$ as their value, and iii) the maximum number of final keys that could be the same as  $\bold{X}^{\eta}_{(q,l)}$ is $\eta_{\text{max}} \left(\frac{U(U-1)}{2}\right)-1$.

Substituting the bounds in  \eqref{Pro-upper-I00I} and \eqref{Pro-upper-IV-II-2} into \eqref{th2Pro-upper-I-0} completes the proof of  Proposition~\ref{Th-exp-pro-2}.
\end{proof}

Finally, substituting the bound in \eqref{maxrta} into \eqref{multi-m-n-p} completes the proof of Theorem~\ref{SemanCrol}.

\subsection{Proof of Theorem~\ref{ch-pl-theor}}\label{Security of the protocol with Public Random Stings}
{According to the protocol, since the final keys $\bold{X}_a$ and $\bold{X}_{(q,l)}$ are computed by using different bits of the binary matrix $\boldsymbol\alpha$ (see \eqref{eq2ql}), they are independent.} Thus, having the plaintext--ciphertext pair $(\bold{M}_a,\bold{C}_a)$ and the binary matrix $\boldsymbol\alpha$, the only way to decrypt the ciphertext $\bold{C}_{(q,l)}$ is that the adversary performs an exhaustive key search to recover the secret key $\bold{Z}_{(q,l)}$. Then, the adversary uses $\bold{Z}_{(q,l)}$ 
to compute $\bold{X}_{(q,l)}$, and consequentially decrypts the message as $\bold{M}_{(q,l)}=\bold{X}_{(q,l)}\oplus\bold{C}_{(q,l)}$.

\begin{figure}
\centering

\subfigure[Security gain as a function of $U$.]{
\includegraphics[width=0.5\textwidth]{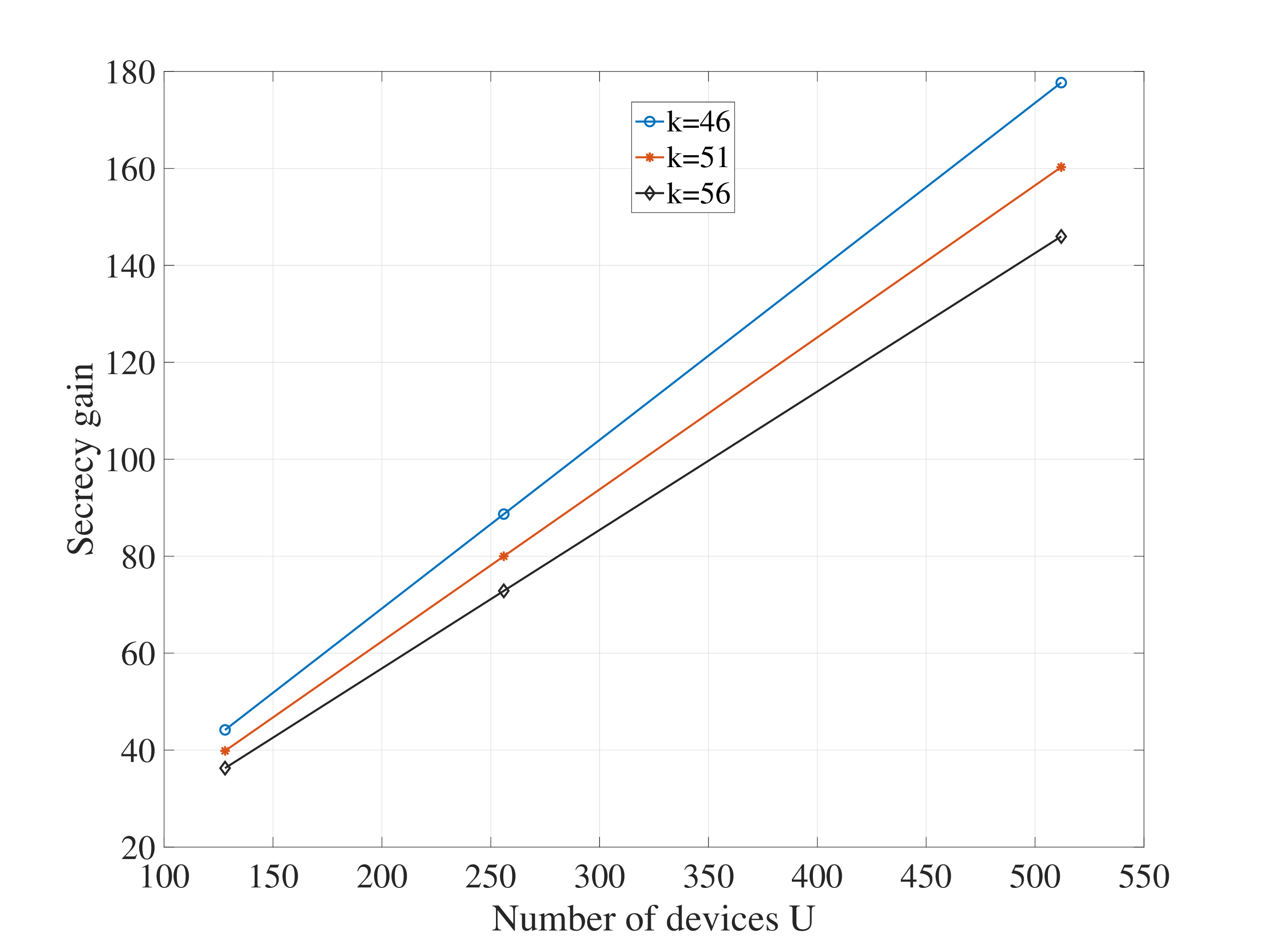}
\label{SGvsU}
}
\subfigure[Advantage of the adversary as a function of $U$.]{
\includegraphics[width=0.5\textwidth]{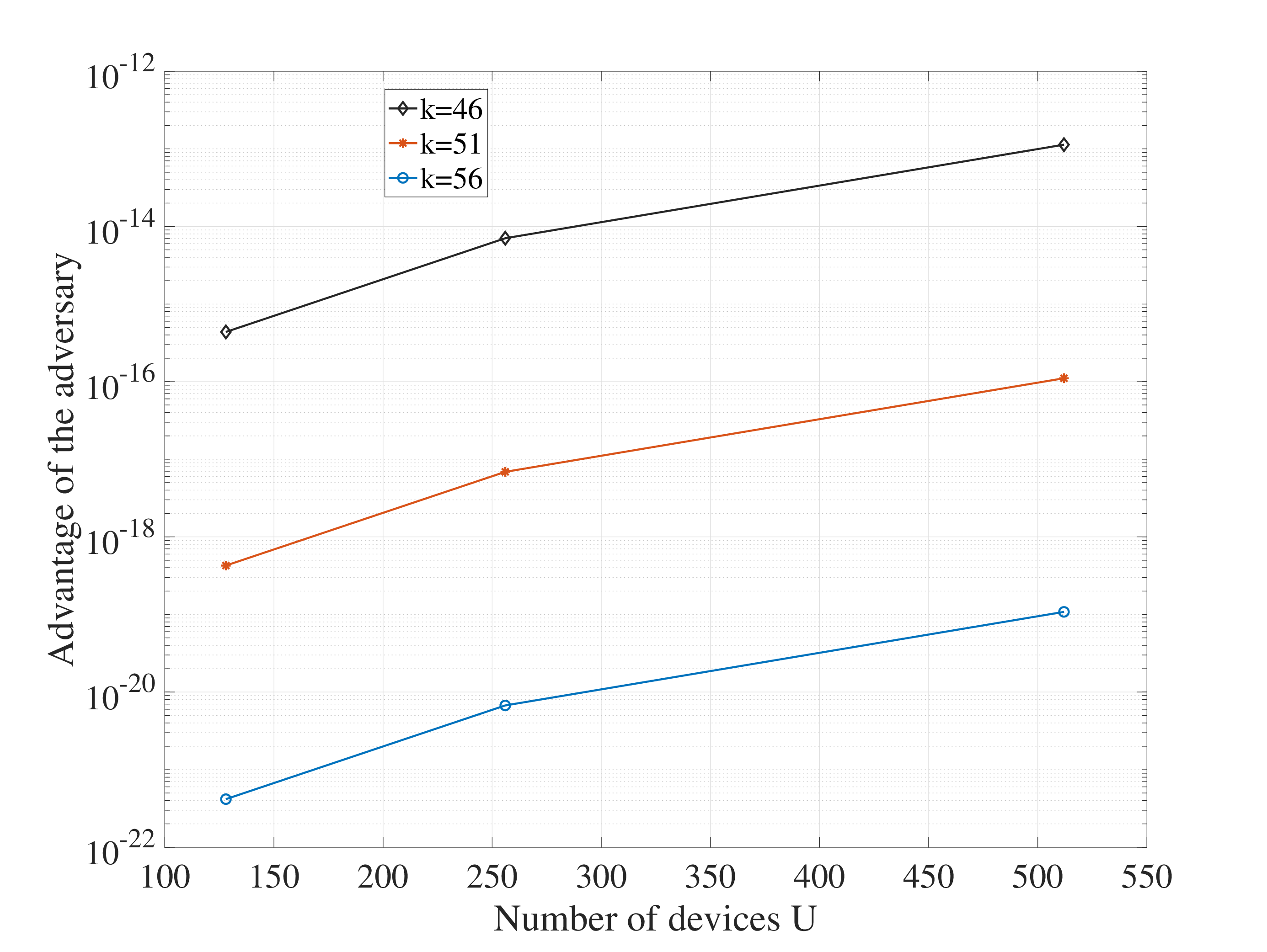}
\label{advvsu}
}
\subfigure[The amount of stored data on each device as a function of $U$.]
{
\includegraphics[width=0.5\textwidth]{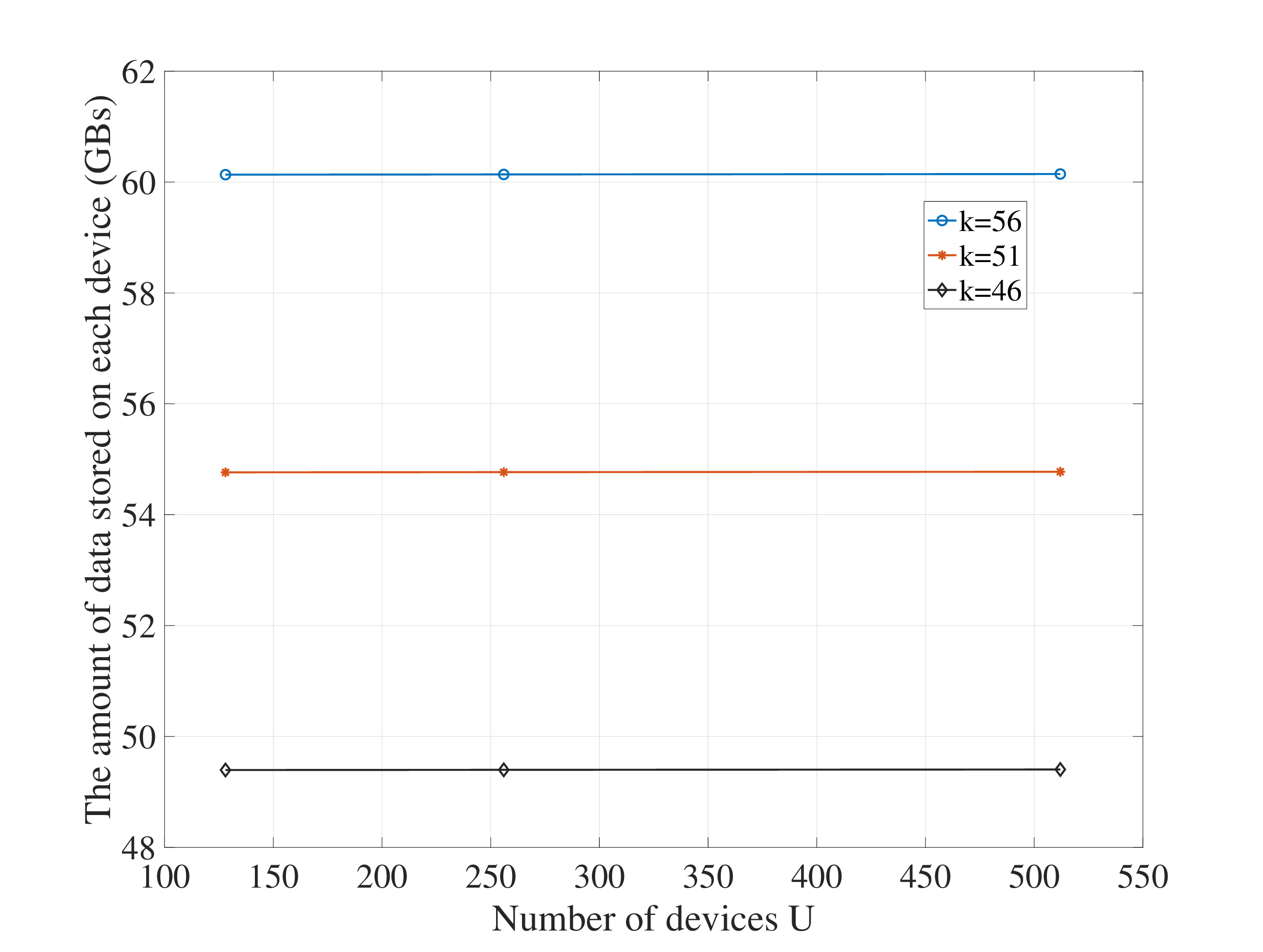}
\label{Storevsu}
}
\caption{Performance of the provided scheme as a function of number of devices $U$ for different values of the security parameter $k$ with  $n=2^{33}$, $\lambda_{(q,l)}=128,$ for all $ (q,l)\in\mathcal{Q}$, $m=2^{10}$, and $\eta_{\text{max}}=\frac{n}{8m}$.}
\label{VSU}
\end{figure}

 First, we show that the probability that the adversary can find the secret key $\bold{Z}_{(q,l)}$ via searching over all possible keys is greater than $1-\frac{1}{2^{m}}$. This is equivalent to showing that the probability that there exists a secret key $\bold{Z}'_{(q,l)}$ with $\bold{Z}'_{(q,l)}\ne\bold{Z}_{(q,l)}$, that provides the same final key as $\bold{X}_a$ is less than $\frac{1}{2^m}$. Let $\bold{X}'_{a}$ denote the final key computed by using the secret key $\bold{Z}'_{(q,l)}$ with $\bold{Z}'_{(q,l)}\ne\bold{Z}_{(q,l)}$. 
    Then, we have
    \begin{align}\nonumber
        &\mathrm{Pr}(\bold{X}_a=\bold{X}'_{a} |\bold{Z}'_{(q,l)}\ne\bold{Z}_{(q,l)})=\\&\nonumber\!\!
      \sum_{\bold{z}_{(q,l)}\in\{0,\ldots,n-1\}^k}\sum_{\bold{{z'}}_{(q,l)}\in\{0,\ldots,n-1\}^k\setminus\{\bold{z}_{(q,l)}\}}\\&\nonumber\mathrm{Pr}(\bold{Z}_{(q,l)}=\bold{z}_{(q,l)}, \bold{Z'}_{(q,l)}=\bold{z'}_{(q,l)})\\& \nonumber\mathrm{Pr}(\bold{X}_a=\bold{X}'_{a} |\bold{Z}'_{(q,l)}\ne\bold{Z}_{(q,l)},\bold{Z}_{(q,l)}=\bold{z}_{(q,l)}, \bold{Z'}_{(q,l)}=\bold{z'}_{(q,l)})\\&
      \stackrel{(a)}{=}\dfrac{n^k(n^k-1)}{n^{2k}2^m}<\dfrac{1}{2^m},
    \end{align}
    where $(a)$ follows because i) $\mathrm{Pr}(\bold{X}_a=\bold{X}'_{a} |\bold{Z}'_{(q,l)}\ne\bold{Z}_{(q,l)},\bold{Z}_{(q,l)}=\bold{z}_{(q,l)}, \bold{Z'}_{(q,l)}=\bold{z'}_{(q,l)})=\frac{1}{2^m}$, and ii) since the pair-wise secret keys $\bold{Z}_{(q,l)}$ and $\bold{Z'}_{(q,l)}$ are independent and uniformly distributed we have $\mathrm{Pr}(\bold{Z}_{(q,l)}=\bold{z}_{(q,l)}, \bold{Z'}_{(q,l)}=\bold{z'}_{(q,l)})=\frac{1}{n^{2k}}$. Thus, the probability that the adversary can find the secret key $\bold{Z}_{(q,l)}$ via searching over all possible keys is greater than $1-\frac{1}{2^{m}}$. 
On the other hand, the size of the secret key is $|\bold{Z}_{(q,l)}|=k\log n$ bits, thus, the adversary needs to test on average $2^{k\log n -1}$ keys to recover $\bold{Z}_{(q,l)}$.

\section{Discussion }\label{Discussion}
In this section, we first present numerical examples and comparative analyses to evaluate the proposed scheme. We then explore potential directions for future research.

\subsection{Numerical Examples}
\begin{figure*}
\centering
\subfigure[The secrecy gain of each device as a function of $n$.]
{
\includegraphics[width=0.48\textwidth]{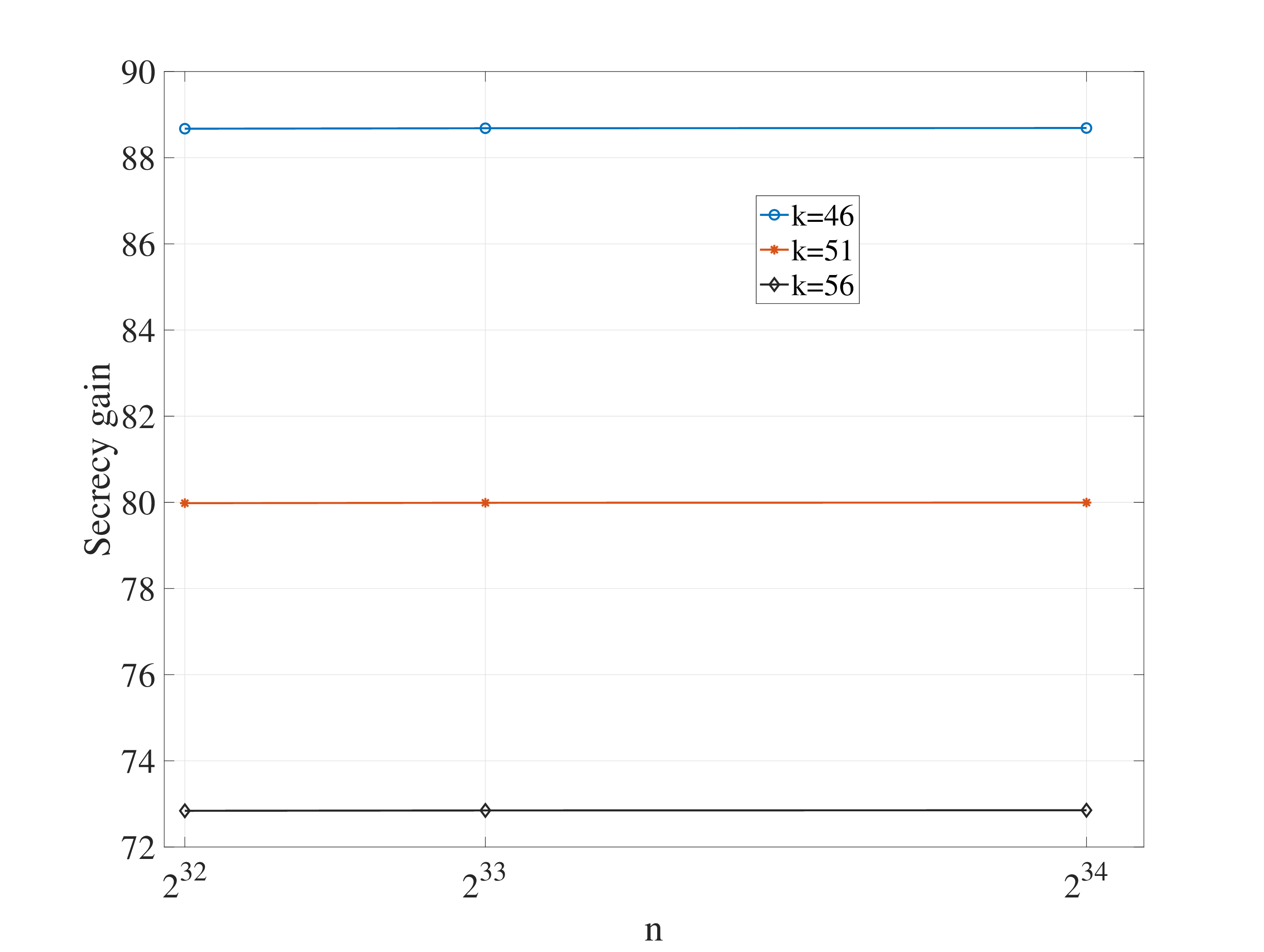}
\label{SGvsN}
}
\subfigure[Advantage of the adversary as a function of $n$.]{
\includegraphics[width=0.48\textwidth]{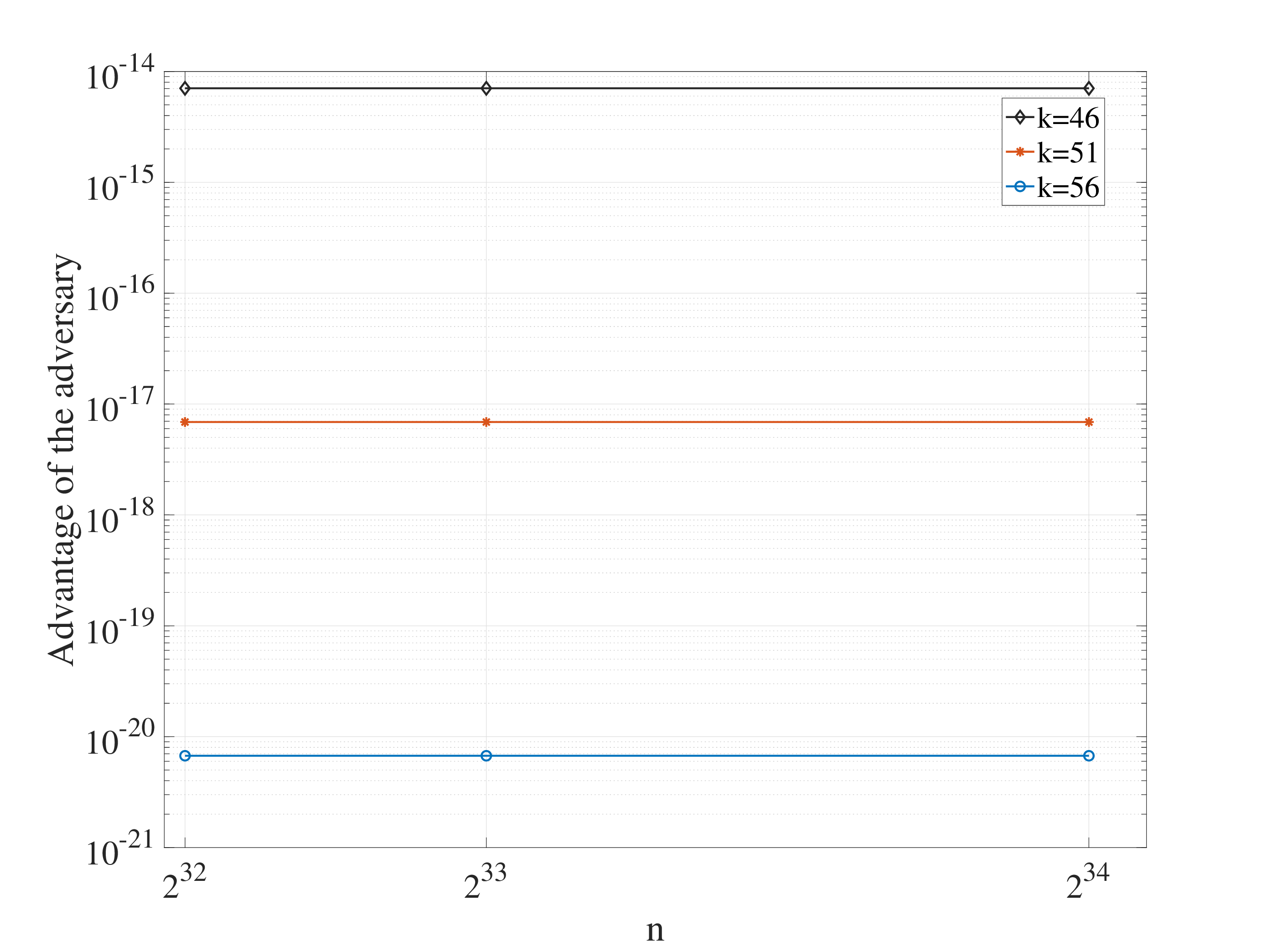}
\label{advvsn}
}
\subfigure[The amount of data that each pair of devices can securely communicate as a function of $n$.]
{
\includegraphics[width=0.48\textwidth]{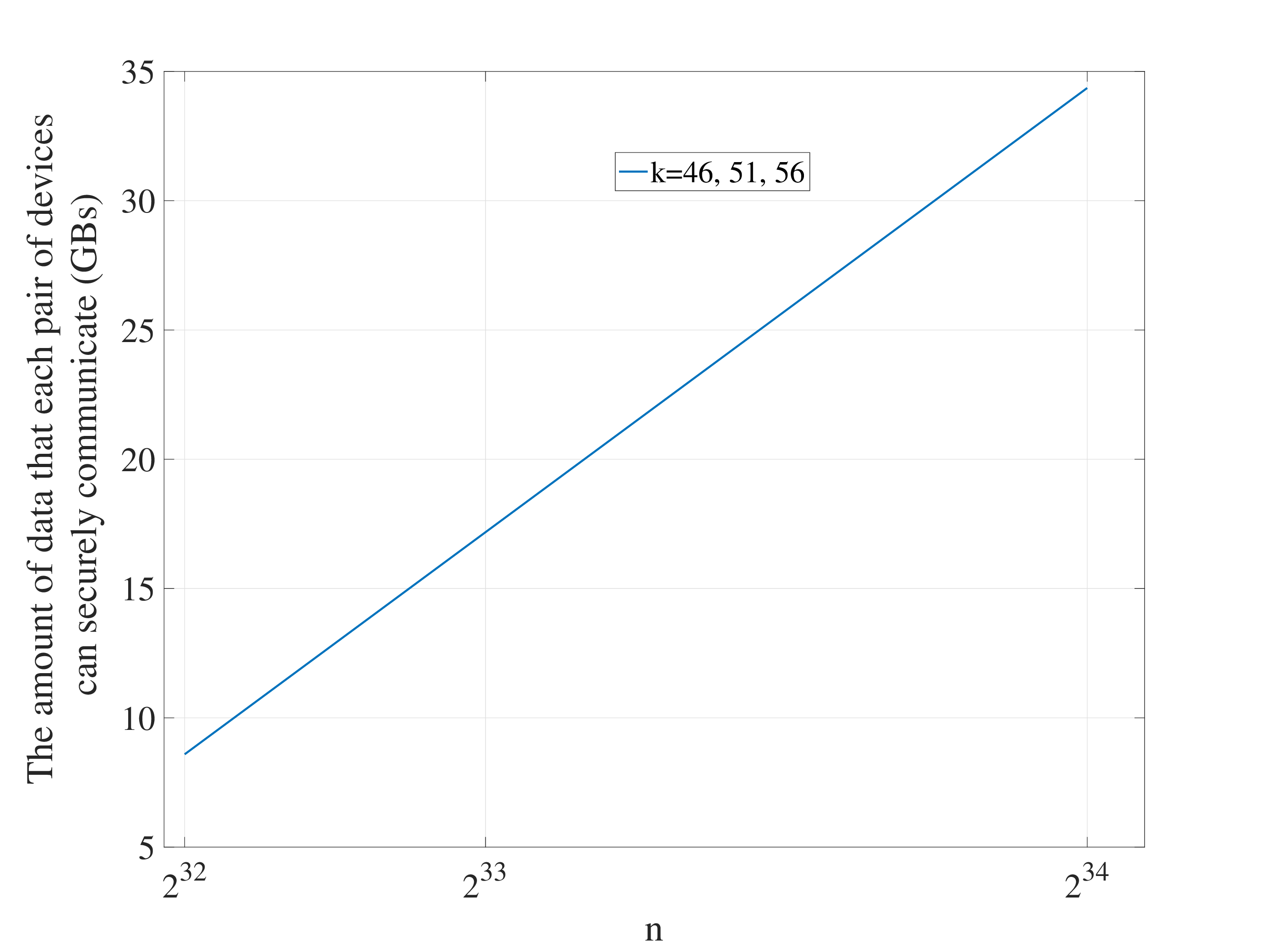}
\label{PerCommunicatevsN}
}
 \subfigure[The amount of stored data on each device as a function of $n$.]
 {
 \includegraphics[width=0.48\textwidth]{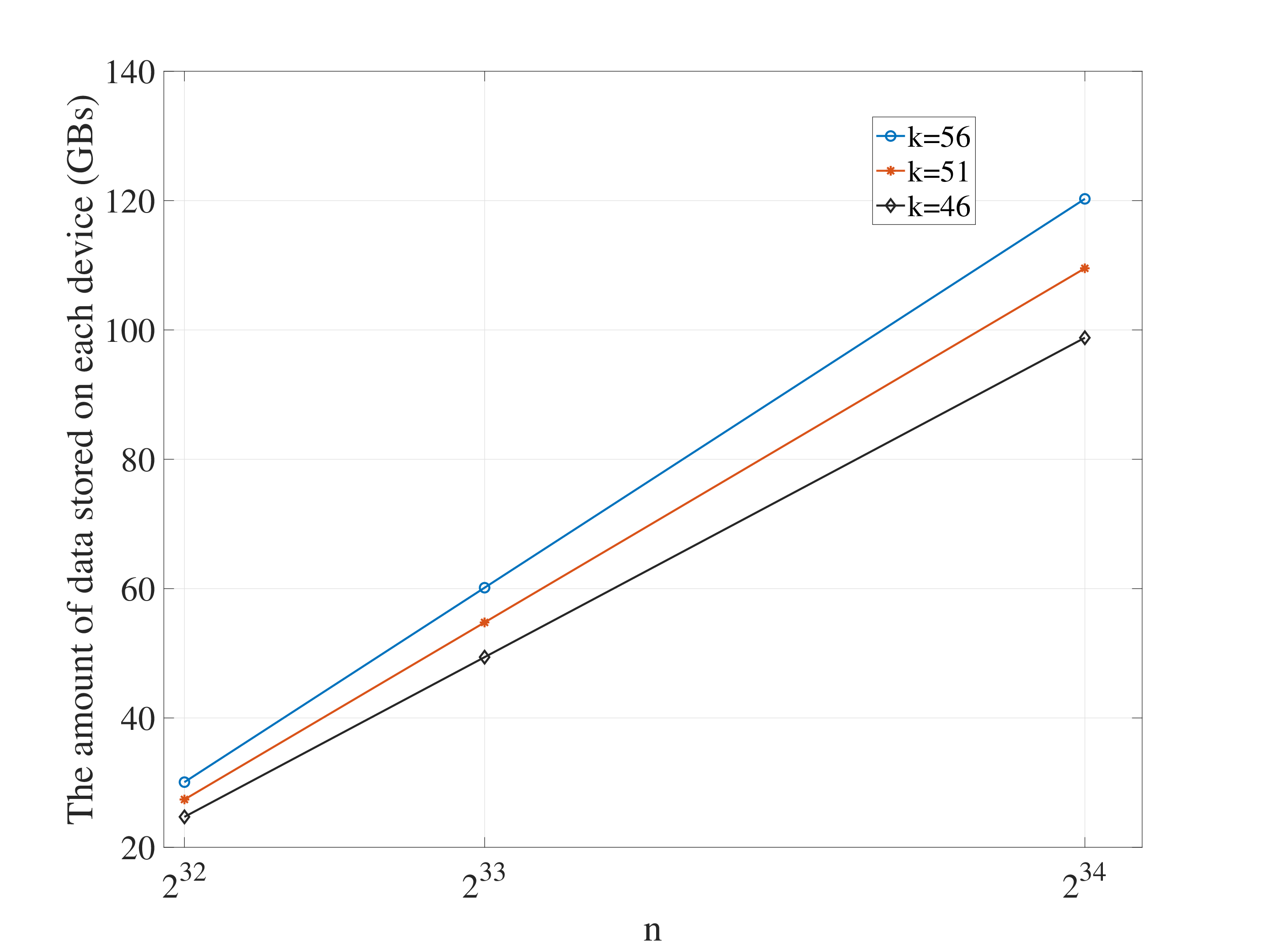}
 \label{Storevsn}
 }
\caption{Performance of the provided scheme as a function of $n$ for different values of the security parameter $k$ with $U=256$, $\lambda_{(q,l)}=128,$ for all $ (q,l)\in\mathcal{Q}$, $m=2^{10}$, and $\eta_{\text{max}}=\frac{n}{8m}$.}
\label{VSN}
\end{figure*}

{We consider an IoBT system consisting of $U$ devices that want to communicate with each other during a mission in the presence of an adversary with unbounded computation power. Fig.~\ref{VSU} illustrates the secrecy gain of each device (in  Fig.~\ref{SGvsU}), the advantage of the adversary (in Fig.~\ref{advvsu}), and the amount of stored data on each device (in Fig.~\ref{Storevsu}) as a function of the number of devices $U$ for different values of the security parameter $k$ with  $n=2^{33}$, number of established pair-wise secret keys between each pair $\lambda_{(q,l)}=128,$ for all $ (q,l)\in\mathcal{Q}$, message length $m=2^{10}$, and maximum number of messages communicated for each established secret key $\eta_{\text{max}}=\frac{n}{8m}$.
%
%
%
As can be seen in Fig.~\ref{SGvsU}, the secrecy gain increases as the number of devices increases and/or the security parameter $k$ decreases. However, one cannot arbitrarily increase the number of devices and/or decrease the security parameter $k$. This is because, as can be seen in Fig.~\ref{advvsu}, the advantage of the adversary is an increasing function of the number of devices and a decreasing function of the security parameter $k$. Thus, there is a trade-off between the secrecy gain of devices and the security guarantee in the system. However, since the advantage of the adversary is a polynomial function of the number of devices $U$ with degree four and an exponentially decreasing function of the security parameter $k$ (see  Theorem~\ref{SemanCrol}), by choosing an appropriate security parameter $k$ one can significantly increase the number of devices, which in turn, increases the security gain. Moreover, Fig.~\ref{Storevsu} shows that by increasing the number of devices, the amount of data required to be stored on each device slightly increases. This is because the size of pair-wise secret keys that are required to be stored for each new device is very small. 
Fig.~\ref{VSN} illustrates the secrecy gain of each device (in Fig.~\ref{SGvsN}), the advantage of the adversary (in Fig.~\ref{advvsu}),  the amount of data that each pair of devices can securely communicate (in Fig.~\ref{PerCommunicatevsN}), and the amount of stored data on each device (in Fig.~\ref{Storevsn}) as a function of $n$ for different values of the security parameter $k$ with the number of devices $U=256$, $\lambda_{(q,l)}=128,$ for all $ (q,l)\in\mathcal{Q}$, $m=2^{10}$, and $\eta_{\text{max}}=\frac{n}{8m}$. 
Figs.~\ref{SGvsN} and \ref{advvsn} show that the secrecy gain and the advantage of the adversary are almost fixed functions of $n$ for the considered parameters. Thus, as shown in Figs.~\ref{PerCommunicatevsN} and \ref{Storevsn}, by storing more data on the devices, one can significantly increase the amount of data that can be exchanged between each pair of devices in the system.}

\subsection{ Comparative Analysis}

\textbf{Evaluation of the avalanche effect:} The avalanche effect measures how sensitive an encryption scheme is to tiny changes in the plaintext. Ideally, flipping a single bit should affect about half of the output bits, ensuring strong randomness in the ciphertext and making it hard for attackers to infer the input. A well‑designed cipher typically achieves close to $50\%$ bit change, which is considered a critical security property.
Table \ref{avalanche} compares the avalanche effect of the proposed scheme versus AES‑128 using plaintexts with varying percentages of zeros, with parameters $n=2^{25}$, $k=30$, and $m=100$ KB. We measure the effect by flipping the first bit of the plaintext. Across all tested input patterns, the proposed scheme consistently exhibits a stronger avalanche effect than AES‑128.


\begin{table*}
\centering
\caption{Avalanche effect evaluation}
\label{avalanche}
\begin{tabular}{ccc}
\toprule
\thead{\% Zeros in Plaintext}
& \thead{Avalanche Effect (Proposed Scheme)}
& \thead{Avalanche Effect (AES‑128)} \\
\midrule
1\%   & 
49.999\% & 50.025\% \\
50\%  &
49.997\% & 49.910\% \\
99\%  & 
50.000\% & 50.052\% \\
\bottomrule
\end{tabular}
\end{table*}

\textbf{Runtime evaluation:} We compare the encryption and decryption runtime of the provided scheme with AES-128. 
 All performance experiments were conducted on a MacBook Pro (Apple M2, 8 GB RAM). The implementation was written in Python using Jupyter Notebook with the following libraries:
  \texttt{pycryptodome} (Crypto.Cipher.AES, Crypto.Random, and Crypto.Util.Padding),  \texttt{numpy}, and \texttt{time} module for timing.
 We used Jupyter’s \texttt{time.perf\_counter()} to measure runtime over multiple runs. Table~\ref{Runtime02} presents the encryption and decryption runtime of the provided scheme and AES-128 for various message sizes and security parameters.


\begin{table*}
\centering
\caption{Runtime comparison}
\label{Runtime02}
\begin{tabular}{lcccc}
\toprule
\thead{Scheme}
& \thead{Message size}
& \thead{Security parameter} 
& \thead{Encryption time}
& \thead{Decryption time}\\
\midrule
Proposed scheme   & $5$ Kb  & $k=10$ & $0.2921$ ms & $0.2921$ ms\\
Proposed scheme   & $5$ Kb  & $k=13$ & $0.3780$ ms & $0.3780$ ms\\
Proposed scheme   & $10$ Kb & $k=10$ & $0.5712$ ms & $0.5712$ ms \\
Proposed scheme   & $10$ Kb & $k=13$& $0.7446$ ms & $0.7446$ ms\\
AES-128           & $5$ Kb  & - & $0.0029$ ms& $0.0030$ ms\\
AES-128           & $10$ Kb & - & $0.0048$ ms& $0.0046$ ms\\
\bottomrule
\end{tabular}
\end{table*}

In addition, in Table~\ref{comparision_table}, we present a comparative analysis of the proposed scheme against the one-time pad and AES-128 schemes from several perspectives, including security, complexity, required secret bits, and storage requirements.

The primary advantage of the proposed scheme over AES-128 lies in its security guarantee: while AES-128 offers computational security, the proposed scheme ensures unconditional security.

Compared to the one-time pad scheme, the proposed scheme is more efficient in terms of the number of secret bits required per encrypted bit. Specifically, for the proposed scheme, this value is $\frac{1}{\Gamma_q}$ for device $q$, whereas it is $1$ for the one-time pad. For instance, with the parameters outlined in Example~1, the required secret bits per encrypted bit is approximately $0.01$ per device.

As shown in Table~\ref{comparision_table}, the complexity of the proposed scheme depends on the security parameter $k$, and while it may be higher than that of the other two schemes. However, this increased complexity primarily involves additional XOR operations, which are generally straightforward to implement and have very low computational overhead.
For example, with parameters in Example~1, the required number of XORs per encrypted bit is $47$.

Although the proposed scheme requires more storage than the other two, the amount remains modest and affordable in practice, typically, on the order of a few GB. For example, with the parameters from Example~1, a device can securely exchange approximately $4381$ GB of data while storing around $17$ GB.

\begin{table*}[t]
    \centering
    \caption{{Competition of the proposed scheme with the one-time pad and AES-128 schemes}}
  {\small  \begin{tabular}{|p{2.5cm}|p{2.6cm}|p{3cm}|p{2.8cm}|p{3.8cm}|}
        \hline
        \textbf{Scheme} & \textbf{Security guarantee} & \textbf{Complexity (number of operations per encrypted bit)} & \textbf{Required secret bits per encrypted bit}& \textbf{Required storage on devices}\\
        \hline
        \textbf{Provided scheme} & Unconditional security & $k+1$ XORs   & $\dfrac{1}{\Gamma_q}$ for device $q$ & $\sum_{l\in\mathcal{U}\setminus \{q\}}\lambda_{(q,l)}k\log n +kn$ \\
        \hline
        \textbf{One-time pad} & Perfect security & $1$ XOR & $1$ & Message size to store the key\\
        \hline
        \textbf{AES-128} & Computational security & Around $43$ XORs and $53$ table look-ups \cite{8422283} & $\dfrac{1}{2^{32.5}}$, \small{\cite[Sect. 4.5]{Boneh2020AppliedCryptography}} & $316$  Bytes \cite{babu2013optimization} \\
        \hline
    \end{tabular}}
    \label{comparision_table}
    
\end{table*}

\subsection{ Dynamic Pair-wise Secret Key Establishment Scheme}

In the proposed protocol, each device is required to establish pair-wise secret keys with all other devices it may need to communicate with during the mission. Storing all these pair-wise keys locally consumes considerable memory and restricts the device from communicating with new devices that do not already share a key. To address this, we propose a \textit{dynamic}, provably secure pair-wise secret key establishment scheme that significantly improves the protocol's flexibility and scalability.

Specifically, a dynamic scheme allows any two devices to establish a pair-wise secret key on demand during the mission, and also enables new devices to join the network at any time.

To realize this, we introduce a designated key distributor in the network. The key distributor is a predetermined device with which all other devices must establish a pair-wise key before the mission begins. By securely communicating with the key distributor using the proposed encryption scheme, any two devices can dynamically establish a pair-wise secret key when needed.

This approach mitigates memory usage on individual devices and supports the addition of new devices by reserving a small number of initial keys at the key distributor for future new devices. However, an important consideration in this setup is the potential failure of the key distributor during the mission. One possible solution is to deploy multiple key distributors in the network to enhance system resilience.

 \subsection{Provably Secure Message Authentication Scheme}

In this paper, we have provided a protocol that addresses the issue of message confidentiality. However, message authenticity and integrity remain crucial aspects of secure communication. Our next goal is to design a message authentication code (MAC) that guarantees authenticity and integrity. This MAC protocol should be provably secure even against adversaries with unbounded computational power. By incorporating mechanisms to ensure authenticity and integrity, a comprehensive and practical framework for achieving perfectly secure communications can be realized.

One of the most important classes of MAC schemes that offer information-theoretic security is universal hashing-based authentication \cite{WChas1981}. Given the high security gain of the proposed protocol, we believe that a portion of the generated final keys between each pair of devices can be used to construct an information-theoretically secure universal hashing-based authentication scheme.

 \subsection{Other Applications}

We believe that the provided encryption scheme can be used to protect sensitive data against adversaries with unbounded computational power in various practical scenarios. For example, consider organizations, such as research centers, health-related centers, and military centers, that are required to store a massive amount of sensitive data for a considerable period.  Using the cloud storage services offered by providers, e.g.,  Dropbox and Amazon, organizations can eliminate the costs associated with setting up and maintaining their storage servers, covering electric power expenses, and managing physical space. However, it is crucial to note that the
 transmission of confidential data to cloud storage often occurs over potentially insecure internet connections and wireless links. Moreover, there is a concern that the cloud storage provider could potentially act as an eavesdropper, making security one of the main concerns that requires careful consideration in this technology. By proper modifications of the provided scheme, the organization will be able to store a massive amount of highly sensitive data on public cloud storage without concern about the privacy of the data. More importantly, thanks to the high secrecy gain of the scheme, organizations are required to store a very small amount of secret bits compared to the amount of data that they can store in public storage. Preliminary results of using the scheme for cloud storage application were presented in \cite{ISIT2024,Asilomar2024}.



\section{Conclusion}\label{Conclusions}
In this paper, we addressed the problem of unconditional secrecy in {an IoBT system} in the presence of an adversary with unbounded computing power and unbounded storage. More specifically, using a shared binary random matrix among the devices, we proposed an encryption scheme that provides everlasting security against the adversary. Moreover, we showed that the scheme is computationally secure against the key recovery attack even if the random matrix is exposed to an adversary with unbounded storage (but with limited computational power). The numerical examples showed that by choosing an appropriate value for the security parameter $k$, one can significantly increase the number of devices in the system, which in turn, increases the secrecy gain.


\bibliographystyle{IEEEtran}
\bibliography{Bibliography}

\end{document}